\documentclass[iop,12pt]{emulateapj}
\usepackage{apjfonts}
\usepackage{natbib}
\usepackage{amsfonts}
\usepackage{amsmath}
\usepackage{multirow}
\usepackage{enumerate}
\usepackage[usenames]{color}
\newcommand{\mev}{\, \text{MeV}}

\def\sun{_\odot}

\def\Dwa{$\,$\uppercase\expandafter{\romannumeral5}$\,$}

\def\sless{\lower2pt\hbox{$\buildrel {\scriptstyle <}
   \over {\scriptstyle\sim}$}}

\def\sgreat{\lower2pt\hbox{$\buildrel {\scriptstyle >}
   \over {\scriptstyle\sim}$}}
\def\sharpnull#1{}

\newcommand{\code}[1]{\texttt{#1}}

\begin{document}
\slugcomment{Received 2010 October 26; accepted 2011 January 10;
published 2011 March 7, ApJ 730 70}

\title{Black Hole Formation in Failing Core-Collapse Supernovae}

\author{Evan O'Connor\altaffilmark{1} and Christian
  D. Ott\altaffilmark{1}} \altaffiltext{1}{TAPIR, Mailcode 350-17,
  California Institute of Technology, Pasadena, CA 91125,
  evanoc@tapir.caltech.edu, cott@tapir.caltech.edu}

\begin{abstract}
  We present results of a systematic study of failing core-collapse
  supernovae and the formation of stellar-mass black holes
  (BHs). Using our open-source general-relativistic 1.5D code
  \code{GR1D} equipped with a three-species neutrino leakage/heating
  scheme and over 100 presupernova models, we study the effects of the
  choice of nuclear equation of state (EOS), zero-age main sequence
  (ZAMS) mass and metallicity, rotation, and mass-loss prescription on
  BH formation.  We find that the outcome, for a given EOS, can be
  estimated, to first order, by a single parameter, the compactness of
  the stellar core at bounce.  By comparing protoneutron star (PNS)
  structure at the onset of gravitational instability with solutions
  of the Tolman-Oppenheimer-Volkof equations, we find that thermal
  pressure support in the outer PNS core is responsible for raising
  the maximum PNS mass by up to 25\% above the cold NS value. By
  artificially increasing neutrino heating, we find the critical
  neutrino heating efficiency required for exploding a given
  progenitor structure and connect these findings with ZAMS
  conditions, establishing, albeit approximately, for the first time
  based on actual collapse simulations, the mapping between ZAMS
  parameters and the outcome of core collapse. We also study the
  effect of progenitor rotation and find that the dimensionless spin
  of nascent BHs may be robustly limited below $a^* = Jc/GM^2 = 1$ by
  the appearance of nonaxisymmetric rotational instabilities.
\end{abstract}


\section{Introduction}
\label{section:intro}
Massive stars with zero-age main sequence (ZAMS) masses
$M_\mathrm{ZAMS}$ in the range of $8-10\,M_\odot \lesssim\,
M_\mathrm{ZAMS} \lesssim 100-150\,M_\odot$ end their lives with the
gravitationally induced catastrophic collapse of their
electron-degenerate iron core to nuclear densities.  There, the
nuclear equation of state (EOS) stiffens and stabilizes the inner
core, which overshoots its new equilibrium and bounces back, launching
a hydrodynamic shock. The shock initially races through the still
collapsing outer core, but soon stalls and turns into an accretion
shock (at $r \sim 100-200\,$km) due to the dissociation of heavy
nuclei at the shock front and neutrino losses from the postshock
region \citep{bethe:90}.  The shock must be revived to drive a
core-collapse supernova (CCSN) and the precise nature of the
responsible CCSN mechanism has been a topic of intense research for
decades (e.g., \citealt{arnett:66,colgate:66,bethewilson:85,janka:07,
burrows:06,burrows:07b,murphy:08,marek:09,nordhaus:10}, and references
therein).

A neutron star (NS) is left behind by a CCSN that explodes soon after
bounce and successfully unbinds its stellar mantle.  However, a
stellar-mass black hole (BH) may be the outcome: (1) if in a
successful, but perhaps weak, CCSN fallback accretion pushes the
nascent NS over its mass limit, (2) if nuclear phase transitions
during protoneutron star (PNS) cooling occur or if PNS cooling reduces
pressure support in a hyper-massive PNS, or (3) if the CCSN mechanism
lacks efficacy and fails to revive the shock and continued accretion
pushes the PNS over its maximum mass.  In this last channel to a
stellar-mass BH, there is no electromagnetic (EM) signal other than
the disappearance of the original star. It is such ``unnovae''
\citep{kochanek:08}, failing CCSNe, that are the topic of this paper.

In ordinary massive stars that hydrostatically form degenerate iron
cores, BH formation, in any scenario, is never prompt
(e.g.,~\citealt{burrows:88}; \citealt{ott:10c}). It is always preceded
by an extended PNS phase giving rise to copious emission of both
neutrinos (\citealt{burrows:88,beacom:01}) and gravitational waves
(\citealt{ott:09}) until the PNS is engulfed by the BH horizon.  The
EM silence expected in a failed CCSN may be broken after all, if
sufficient and appropriately distributed angular momentum is present
to allow for a Keplerian accretion disk to form near the BH,
permitting a collapsar (\citealt{woosley:93}) gamma-ray burst (GRB)
central engine to operate and drive relativistic outflows.

It is currently unclear what fraction of massive stars form BHs and
through which channel. Preexplosion observations of progenitors of
successful CCSNe suggest progenitor masses $\lesssim 17-20\,M_\odot$
\citep{smartt:09a} for standard Type II-P supernovae. Assuming, as
suggested by \cite{smartt:09a}, that most other CCSNe fail or make BHs
after a successful explosion, this would correspond to a BH fraction
of $\lesssim 30\,$\%-$35\,$\% of massive stars above
$8\,M_\odot$. However, alternative interpretations exist and have been
summarized by \cite{smith:10}.  Theoretical work by \cite{timmes:96},
\cite{fryer:99}, \cite{heger:03}, and \cite{eldridge:04} provided
rough estimates on the outcomes of stellar collapse as a function of
progenitor ZAMS mass and metallicity. Leaving effects due to binary
evolution aside, \cite{zhang:08} performed an extensive study of
fallback in artificially driven spherically symmetric CCSN explosions
and estimated that zero-metallicity stars form BHs in
$20\,$\%-$50\,$\% of all core-collapse events with an average BH mass
of $6-10\,M_\odot$.  For solar-metallicity stars, due to increased
mass loss during evolution, \cite{zhang:08} found BHs to form at a
significantly lower rate and initial mass.  They predict BH fractions
in the range of $10\,$\%-$25\,$\% with typical initial BH masses of
$3\,M_\odot$. This is in rough agreement with previous population
synthesis calculations of \cite{fryer:01} and \cite{belczynski:02}.

Early spherically symmetric (one-dimensional, 1D) simulations of BH
formation in failing CCSNe were carried out by \cite{wilson:71} and
\cite{vanriper:78b}.  \cite{burrows:88} performed a set of
quasi-stationary 1D PNS accretion and cooling simulations to
investigate the possibility of BH formation in SN 1987A.  Delayed BH
formation (by tens of seconds), due to, e.g., a nuclear EOS phase
transition, was studied by \cite{baumgarte:96b, baumgarte:96}. More
recently, 1D full Boltzmann neutrino radiation-hydrodynamics
calculations of failing nonrotating CCSNe were carried out by
\cite{liebendoerfer:04} and more recently
\cite{sumiyoshi:07,sumiyoshi:08,sumiyoshi:09} and \cite{fischer:09a}.
These studies provided detailed neutrino signature predictions for
BH-forming core-collapse events. However, owing to the complexity and
computational expense of such Boltzmann-transport calculations, these
groups could consider only very limited sets of progenitor models and
EOS. Simplified axisymmetric (2D) simulations of BH formation in
rotating core collapse were first performed in a series of papers by
\cite{sekiguchi:04,sekiguchi:05} and \cite{shibata:05}. These authors
used simplified EOS, no neutrino treatment, and artificially
constructed initial conditions and found prompt BH
formation. Recently, the same authors performed a small set of 2D
simulations with a finite-temperature nuclear EOS and a leakage scheme
for neutrinos \citep{sekiguchi:10c} and considered collapse, BH
formation, and subsequent evolution in an artificially constructed
progenitor with an iron core mass of $\sim 13\,M_\odot$ and constant
specific entropy of $8\,{k_B}$/baryon, initial conditions that are
inconsistent with those at the precollapse stage of CCSN progenitors.

In this paper, our focus is on studying and establishing the
systematics of failing CCSNe and BH formation.  For this, we employ
the spherically symmetric general-relativistic (GR) open-source code
\code{GR1D} \citep{oconnor:10} that can handle rotation in an
approximate angle-averaged way (``1.5D'') and sacrifice accuracy in
the neutrino treatment by employing an efficient energy-averaged
three-species neutrino leakage/heating scheme instead of full
transport.  The efficiency of \code{GR1D} enables us to perform more
than $\sim 700$ collapse calculations, investigating for the first
time in detail the effects of variations in nuclear EOS, progenitor
ZAMS mass and metallicity, neutrino heating efficiency, and
precollapse rotational configuration. We employ four different
finite-temperature nuclear EOS and draw a total of 106 progenitor
models from seven stellar evolution studies.

In Section \ref{sec:methods}, we review the features of our 1.5D GR
hydrodynamics code \code{GR1D}, discuss our neutrino leakage/heating
scheme, and introduce the set of employed
EOS. Section~\ref{sec:initialmodels} introduces our progenitor model
set, numerical grid setup, and precollapse rotational setup. In
Section \ref{sec:fid}, we introduce key aspects of failing CCSNe and BH
formation by discussing the evolution of BH formation in a fiducial
nonrotating $40\,M_\odot$ solar-metallicity progenitor. We go on in
Section \ref{sec:eosdependence} to study the influence of the EOS and
thermal effects on the time to BH formation and on the maximum
(baryonic and gravitational) mass of the PNS. We discover that for
nuclear EOS with physically plausible stiffness, the maximum (baryonic
and gravitational) mass of the PNS is always greater than the
corresponding cold NS mass and discuss that the difference is due
entirely to thermal pressure support of material in the hot outer PNS
core. This effect is strongest for the softest considered EOS and
decreases with increasing EOS stiffness. In Section \ref{sec:prestruct}, we
analyze the impact of variations in progenitor structure on the time
to BH formation and the maximum PNS mass in failing CCSNe. We find
that the postbounce dynamics can be predicted rather robustly by a
single parameter, the compactness of the progenitor structure
at core bounce. The same approximate single-parameter dependence
emerges in Section \ref{sec:lums}, where we determine the neutrino heating
efficiencies required (modulo ignored multi-dimensional effects) to induce a
neutrino-driven explosion in a large set of progenitors. The combined
results of Sections \ref{sec:prestruct} and \ref{sec:lums} allow us to make
predictions on the outcome of core collapse for progenitors with
varying ZAMS mass and metallicity in Section \ref{sec:evolution}. As we
discuss in that section, mass loss may be the greatest uncertainty in
connecting ZAMS parameters to core-collapse results. In
Section \ref{sec:rotatingBH}, we present results from the first rotating BH
formation simulations in the CCSN context. Varying the precollapse
rotation rate in Section \ref{sec:parameterizedrotation}, we find that, not
unexpectedly, increased rotation leads to a delay of BH formation and
greater maximum PNS masses. We also observe that the birth spin of
Kerr BHs in nature appears to be robustly limited to values below
$a^\star = J/M^2 \lesssim 0.9$ by the likely appearance of
nonaxisymmetric rotational instabilities that redistribute or radiate
angular momentum.  This finding requires confirmation by
3D simulations.  We go on in Section \ref{sec:grb} to discuss the collapse
evolution of a set of progenitors that were evolved from the ZAMS with
a 1.5D treatment of rotation and discuss their viability as
collapsar-type long-GRB progenitors. Finally, in
Section \ref{sec:conclusion}, we critically summarize our work and conclude.

\section{Methods}
\label{sec:methods}

\subsection{GR Hydrodynamics}
\code{GR1D} \citep{oconnor:10} is a spherically symmetric GR
hydrodynamics code developed for the study of stellar collapse and BH
formation.  It is available for download at {\tt
  http://stellarcollapse.org}.  \code{GR1D}, based on the previous
work of \cite{gourgoulhon:91} and \cite{romero:96}, is Eulerian and
uses the radial gauge -- polar slicing coordinates that have the
simplifying property of a vanishing shift vector.  Here, we briefly
outline the basics of the curvature and hydrodynamics equations and refer
the reader to \cite{oconnor:10} for full details and derivation. We
assume spacelike signature $(-,+,+,+)$ and, unless noted otherwise,
use units of $G = c = M_\odot = 1$. The metric of \code{GR1D} is given
by the line element
 \begin{eqnarray}
 ds^2 & = & -\alpha(r,t)^2 dt^2 + X(r,t)^2 dr^2 + r^2 d\Omega^2\,,
\end{eqnarray}
where $\alpha (r,t) = \exp(\Phi(r,t))$ with $\Phi(r,t)$ being the
metric potential and $X(r,t) = [1-2m(r)/r]^{-1/2}$, where $m(r)$
is the enclosed gravitational mass.  We assume an ideal fluid 
with stress energy given by
\begin{equation}
T^{\mu \nu} = \rho h u^\mu u^\nu + g^{\mu \nu} P\,,\label{eq:SET}
\end{equation}
where $\rho$ is the matter density, $P$ is the fluid pressure and $h =
1+\epsilon+P/\rho$ is the specific enthalpy with $\epsilon$ being the
specific internal energy.  $u^\mu$ in Equation~(\ref{eq:SET}) is the
4-velocity of the fluid, and without rotation, taken to be $u^\mu =
(W/\alpha,Wv^r,0,0)$, where $W = [1 - v^2]^{-1/2}$ is the Lorentz
factor and $v = Xv^r$ is the physical velocity.  For a given matter
configuration, the Hamiltonian and momentum constraint equations give
differential equations for both $m(r)$ and $\Phi(r)$,
\begin{equation}
  m(r) = 4\pi \int_0^r (\rho h W^2 - P + \tau_m^\nu) {r^\prime}^2 dr^\prime\,,\label{eq:mass}
\end{equation}
\begin{equation}
  \Phi(r,t) = \int_0^rX^2\left[{m(r^\prime,t) \over {r^\prime}^2} +
    4\pi r^\prime(\rho h X^2
    u^{r^\prime}u^{r^\prime} + P + \tau_\Phi^\nu)\right]dr^\prime + \Phi_0\,\,,\label{eq:phi}
\end{equation}
where $\Phi_0$ is determined by matching the metric at the star's
surface to the Schwarzschild metric. The neutrino terms,
$\tau_m^\nu$ and $\tau_\Phi^\nu$, account for trapped neutrinos and
their detailed form is given in \cite{oconnor:10}.  We obtain the
fluid evolution equations by expanding the local fluid rest-frame
conservation laws, $\nabla_\mu T^{\mu \nu} = 0$ and $\nabla_\mu J^\mu
= 0$, in the coordinates of \code{GR1D}. The conservation laws
become
\begin{equation}
\partial_t\left(\vec{U}\right) + {1 \over r^2}\partial_r\left({\alpha r^2 \over
    X} \vec{F}\right) = \vec{\cal{S}}\,\,,\label{eq:evolution}
\end{equation}
where $\vec{U} = \left(D,\ DY_e,\ S^r,\ \tau\right)$ are the conserved
variables,  given in terms of the primitive fluid
variables $\rho,\ Y_e,\ \epsilon,\ P \mathrm{,\,and}\ v$ as
\begin{equation}
\vec{U} = \left(\begin{array}{c}
D\\
DY_e\\
S^r\\
\tau\end{array}\right) =
\left(\begin{array}{c}
X\rho W\\
X\rho WY_e\\
\rho h W^2 v\\
\rho h W^2 - P - D
\end{array}\right)\,.
\end{equation}

The spatial fluxes in Equation~(\ref{eq:evolution}) are given by,
\begin{equation}
\vec{F} = (Dv,\ DY_ev,\ S^rv+P,\ S^r-Dv)\,,
\end{equation} 
and the source terms
\begin{eqnarray}
  \vec{\mathcal{S}} =& \bigg[ 0 ,\ R^\nu_{Y_e} ,\ (S^rv -
  \tau - D)\alpha X\left(8 \pi r P + \frac{m}{r^2}\right) + \alpha P X
  {m \over r^2}  \nonumber \\
  &+ {2 \alpha P \over X r} + Q_{S^r}^{\nu,\mathrm{E}} +
  Q_{S^r}^{\nu,\mathrm{M}},\ Q_\tau^{\nu,\mathrm{E}} + Q_\tau^{\nu,\mathrm{M}}
  \bigg]\,,
\label{eq:sources}
\end{eqnarray}
where the $R$s and $Q$s are neutrino sources and sinks which arise
from the neutrino leakage scheme and neutrino pressure
contributions (see~\citealt{oconnor:10} for details).

The evolution equations (Equation~\ref{eq:evolution}) are first spatially
discritized using a finite-volume scheme (e.g.,
\citealt{romero:96,font:08}).  The piecewise parabolic method (\citealt{colella:84}) is used to reconstruct the state variables at the
cell interfaces and the HLLE Riemann solver \citep{HLLE:88} is
employed to determine the physical fluxes through these interfaces.
The evolution equations are integrated forward in time via the method
of lines \citep{Hyman-1976-Courant-MOL-report}, using standard
second order Runge-Kutta time integration with a Courant factor of
$0.5$.  After updating the conserved variables, they are inverted via
a Newton-Raphson scheme to obtain the new fluid state variables.

In spherical symmetry, rotation can be included by assuming constant
angular velocity $\Omega$ on spherical shells (shellular rotation) and
including an angle-averaged centrifugal force in the radial momentum
equation. This is common practice in stellar evolution codes (e.g.,
\citealt{heger:00}) an has also been applied to Newtonian 1D stellar
collapse calculations~\citep{akiyama:03,thompson:05}. We include a GR
variant of this ``1.5D'' rotation treatment in \code{GR1D} (1) by
adding an evolution equation for the generalized specific angular
momentum $S_\phi = \rho h W^2 v_\varphi r$, (2) by including an
effective centrifugal force in the equation for $S^r$, and (3) by
modifying the expressions for the 4-velocity, the Lorentz factor, and
the differential equation for the metric potential to account for
rotation. Full details as well as a demonstration of conservation of
angular momentum can be found in \cite{oconnor:10}. Note that, as may
be expected and was demonstrated by \cite{ott:06spin}, the 1.5D
approximation becomes less accurate with increasing spin and
quantitative results are reliable only for low rotation rates.

\subsection{Neutrino Treatment}
\label{sec:neutrino}

Neutrino effects are crucial in stellar collapse and should ideally be
included via a computationally expensive GR Boltzmann transport
treatment (e.g., \citealt{liebendoerfer:04}). However, since our aim
is to perform an extensive parameter study with hundreds of
simulations, we choose to resort to a less accurate, but much more
computationally efficient leakage and approximate heating scheme for
neutrinos. Details of this are laid out in \cite{oconnor:10}. Here we
review only its most salient features.

Before core bounce, neutrinos deleptonize the collapsing core,
reducing the electron fraction $Y_e$ and, as a consequence, the size
of the homologous inner core \citep{bethe:90}.
\cite{liebendoerfer:05fakenu} showed that the prebounce $Y_e$ can be
parameterized as a function of density and that this parameterization
varies little between progenitor stars.  We follow this prescription
for prebounce deleptonization and use the $Y_e(\rho)$ fit parameters
of his G15 model.  Following bounce, this simple parameterization
becomes inaccurate and cannot capture the effects of neutrino cooling,
deleptonization, and neutrino heating.  Hence, we switch to a leakage
scheme that uses elements of what was laid out by \cite{ruffert:96}
and \cite{rosswog:03b}. We consider three neutrino species, $\nu_e$,
$\bar{\nu}_e$, and $\nu_x =
\{\nu_\mu,\bar{\nu}_\mu,\nu_\tau,\bar{\nu}_\tau\}$.  Neutrino pairs of
all species are made in thermal processes of which we include
electron-positron pair annihilation and plasmon decay
\citep{ruffert:96}.  In addition, charged-current processes lead to
the emission of $\nu_e$s and $\bar{\nu}_e$s. The leakage scheme
provides approximate energy and number emission rates that are
inserted into \code{GR1D}'s evolution equations via source terms in
Equation~(\ref{eq:sources}), $R^\nu_{Y_e},\ Q^{\nu,\mathrm{E}}_{S^r},\
\mathrm{and}\ Q^{\nu,\mathrm{E}}_{\tau}$ \citep{oconnor:10}.

We include neutrino heating via a parameterized charged-current
heating scheme based on \cite{janka:01}. The heating rate at radius $r$ is
\begin{equation}
Q^{\mathrm{heat}}_{\nu_i}(r) = f_\mathrm{heat} \frac{L_{\nu_i}(r)}{4\pi r^2}
\sigma_{\mathrm{heat},\nu_i}\, {\rho\over m_u} X_i \left\langle {1 \over
  F_{\nu_i}} \right\rangle e^{-2\tau_{\nu_i}} \,\,,\label{eq:heating}
\end{equation}
where $f_{\mathrm{heat}}$ is a scale factor that allows for
artificially increased heating, $L_{\nu_i}(r)$ is the neutrino
luminosity interior to $r$, $\tau_{\nu_i}$ is the optical depth,
determined through the leakage scheme, $\sigma_\mathrm{heat,\nu_i}$ is
the energy-averaged absorption cross section, and $X_i$ is
corresponding mass fraction of the neutrino reaction ($X_p$ for
$\bar{\nu}_e$ capture on protons and $X_n$ for
$\nu_e$ capture on neutrons). $\langle 1/F_{\nu_i}\rangle$ is the
mean inverse flux factor which we approximate analytically as a
function of the optical depth $\tau$ by comparing to the
angle-dependent radiation transport calculations of \cite{ott:08}.

We include in our simulations the stabilizing effect of neutrino
pressure in the optically thick PNS core via an ideal Fermi-gas
approximation (\citealt{liebendoerfer:05,oconnor:10}).  Leaving out
this pressure contribution leads to $\sim 5$\% smaller maximum
gravitational PNS masses. We also include terms due to neutrino
pressure and radiation-field energy in the calculation of the
gravitational mass (Equation~\ref{eq:mass}) and of the metric potential
(Equation~\ref{eq:phi}). Since our leakage scheme does not treat neutrino
energy separately from the internal energy of the fluid, including the
energy of the neutrino gas in the former equations is not fully
consistent with our present approach. This error was discovered and
corrected after all simulations were performed. However, a set of test
calculations showed that the error leads to an underestimate of the
maximum gravitational PNS mass of only $\sim 2$\% which is well within
the error of the overall leakage scheme (see also
Section \ref{sec:previousworks}).

\subsection{Equations of State and Maximum Neutron Star Masses}
\label{sec:EOS}

We include multiple finite-temperature nuclear EOS in this study to
explore the dependence of postbounce evolution and BH formation on EOS
properties. The Lattimer-Swesty (LS) EOS \citep{lseos:91} is based on
the compressible liquid-droplet model, assumes a nuclear symmetry
energy $S_v$ of $29.3\,$MeV, and comes in three variants with
different values of the nuclear incompressibility of $K_\mathrm{s}=
180\,$MeV (LS180), $220\,$MeV (LS220), and $375\,$MeV (LS375). The EOS
of \cite{shen:98a,shen:98b} (HShen EOS), on the other hand, is based
on a relativistic mean-field model, has $S_v = 36.9\,$MeV and
$K_\mathrm{s}=281\,$MeV. More details on these EOS and their
implementation in \code{GR1D} is given in \cite{oconnor:10}. The EOS
tables and driver routines employed in this study are available for
download at ${\tt http://stellarcollapse.org}$.

\begin{figure}[t] 
\begin{center} 
\includegraphics[width=\columnwidth]{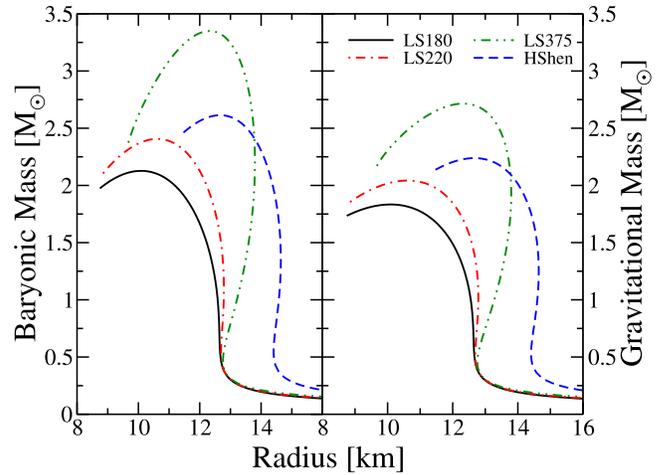} 
\caption{Baryonic (left) and gravitational (right) neutron
  mass--radius relations for various hot nuclear EOS.  The
  temperature is taken to be constant throughout the star at
  $T=0.1\mev$ and the electron fraction is determined through
  neutrinoless $\beta$-equilibrium with an imposed minimum of 0.05
  due to table constraints. }\label{fig:MvsRtemp0.1}
\end{center}
\end{figure}

By solving the Tolman-Oppenheimer-Volkoff (TOV) equations
\citep{oppenheimer:39} with $T=0.1\,$MeV and assuming neutrinoless
$\beta$ equilibrium we determine the neutron star baryonic and
gravitational mass--radius relationships that each of these four EOS
produces and that are depicted by Figure~\ref{fig:MvsRtemp0.1}.  The
maximum gravitational (baryonic) neutron star masses are $\sim 1.83
\,M_\odot$ ($\sim 2.13 \, M_\odot$), $\sim 2.04 \,M_\odot$ ($\sim 2.41
\, M_\odot$), $\sim 2.72 \,M_\odot$ ($\sim 3.35 \, M_\odot$), and
$\sim 2.24 \,M_\odot$ ($\sim 2.61 \,M_\odot$) for LS180, LS220, LS375,
and HShen, respectively.  The coordinate radii of these maximum-mass
stars are $\sim 10.1\,$km, $\sim 10.6\,$km, $\sim 12.3\,$km and $\sim
12.6\,$km, respectively.  

The above maximum neutron star masses hold only for nonrotating cold
NSs.  As we will discuss in detail in Section \ref{sec:eosdependence}, the
PNSs at the heart of the failing CCSNe considered in this work are
much hotter. They have central temperatures of $\sim 10-20\,$MeV and
tens of MeV in their outer core and mantle. Thermal effects
have a significant effect on their maximum masses.

In this study, we do not consider hyperonic EOS, e.g., the hyperonic
extension of the HShen EOS by \cite{ishizuka:08}, or EOS involving
other phases of nuclear matter, e.g., quarks and pions
\cite{nakazato:10}. Such EOS are potentially interesting in failing
CCSNe, since their exotic components lead to a softening of the EOS at
high density, potentially accelerating BH formation
\citep{sumiyoshi:09}. We also do not consider EOS that include QCD
phase transitions that too may lead to early PNS collapse and
potentially to a second bounce and neutrino burst \citep{sagert:09}.

\section{Model Setup}
\label{sec:initialmodels}
\subsection{Presupernova Data}
\label{sec:presuperdata}
We make use of single-star nonrotating presupernova models from
several stellar evolution studies: \citet{ww:95} (WW95),
\citet{whw:02} (WHW02), \citet{limongi:06} (LC06A/B) and
\citet{woosley:07} (WH07).  Each of these studies evolved stars with a
range of ZAMS masses at solar metallicity ($Z_\odot$, hereafter
denoted with prefix $s$ in model names) up until the onset of core
collapse.  In addition to solar metallicity, WHW02 evolved stars with
ultra low metallicity, $10^{-4}\,Z_\odot$ (denoted by prefix $u$) and
zero metallicity (denoted by prefix $z$).  Rotation is of relevance in
stellar evolution and stellar evolutionary processes affect the
rotational configuration at the presupernova stage.  In order to study
BH formation, BH birth properties and their impact on a potential
subsequent evolution to a GRB in such spinning progenitors, we draw
representative models from \cite{heger:00} (HLW00) and from
\cite{woosley:06} (WH06) who included rotation in essentially the same
way as we do in \code{GR1D}.

In Table~\ref{tab:initialmodels}, we list key parameters for all
models in our set.  These include presupernova mass, iron core mass
(which we define as the baryonic mass interior to $Y_e = 0.495$), and
the bounce compactness $\xi_{2.5}$.  The latter is defined as
\begin{equation}
  \xi_{M} = {M / \,M_\odot  \over R(M_\mathrm{bary} =
    M) / 1000\,\mathrm{km}}\Big|_{t =t_{\mathrm{bounce}}}\,,\label{eq:bouncecompactness}
\end{equation}
where we set $M = 2.5\,M_\odot$. $R(M_\mathrm{bary}=2.5\,M_\odot)$ is the
radial coordinate that encloses 2.5-$M_\odot$ at the time of core
bounce.  $\xi_{2.5}$ gives a measure of a progenitor's
compactness at bounce. We choose $M = 2.5\,M_\odot$ as this is the
relevant mass scale for BH formation. $\xi_{2.5}$ is, as we shall
discuss in Section \ref{sec:prestruct}, a dimensionless variable that allows
robust predictions on the postbounce dynamics and the evolution of the
model toward BH formation.  The evaluation of $\xi_{2.5}$ at core
bounce is crucial, since this is the only physical and unambiguous
point in core collapse at which one can define a zero of time and can
describe the true initial conditions for postbounce
evolution. Computing the same quantity at the precollapse stage leads
to ambiguous results, since progenitors come out of stellar evolution
codes in more or less collapsed states. Collapse washes out these
initial conditions and removes ambiguities.

We point out (as is obvious from Table~\ref{tab:initialmodels}) that
there is a clear correlation between iron core mass and bounce
compactness.  Since the effective Chandrasekhar mass
increases due to thermal corrections \citep{burrows:83,baron:90}, more massive
iron cores are hotter. Hence, progenitors with greater bounce
compactness result in higher-temperature PNSs.

\begin{deluxetable}{rcccc}
  \tablecolumns{5}
  \tablewidth{0pc}
  \tablecaption{Initial Models}
   \tablehead{
    \multirow{2}{*}{Model} & $M_{\mathrm{ZAMS}}$ & $M_{\mathrm{pre-SN}}$  &  $M_{\mathrm{Fe\                 
        core}}$\tablenotemark{a}  & $\xi_{2.5}$\tablenotemark{b}\\
    &\multirow{1}{*}{$(M_\odot)$}&\multirow{1}{*}{$(M_\odot)$}&\multirow{1}{*}{$(M_\odot)$}&
 }
\startdata
  s20WW95 & \phantom{0}20 & 20.0\phantom{0} & 1.74 & 0.383\\
  s25WW95 & \phantom{0}25 & 25.0\phantom{0} & 1.77 & 0.416\\
  s40WW95 & \phantom{0}40 & 40.0\phantom{0} & 1.98 & 0.583\\
  \hline
  s15WHW02 & \phantom{0}15 & 12.6\phantom{0} & 1.55 & 0.150\\
  s20WHW02 & \phantom{0}20 & 14.7\phantom{0} & 1.46 & 0.127\\
  s25WHW02 & \phantom{0}25 & 12.5\phantom{0} & 1.62 & 0.326\\
  s30WHW02 & \phantom{0}30 & 12.2\phantom{0} & 1.46 & 0.223\\
  s35WHW02 & \phantom{0}35 & 10.6\phantom{0} & 1.49 & 0.205\\
  s40WHW02 & \phantom{0}40 & \phantom{0}8.75 & 1.56 & 0.266\\
  s75WHW02 & \phantom{0}75 & \phantom{0}6.36 & 1.48 & 0.112\\
  \hline
  u20WHW02 & \phantom{0}20 & 20.0\phantom{0} & 1.57 & 0.338\\
  u25WHW02 & \phantom{0}25 & 25.0\phantom{0} & 1.53 & 0.223\\
  u30WHW02 & \phantom{0}30 & 30.0\phantom{0} & 1.58 & 0.326\\
  u35WHW02 & \phantom{0}35 & 35.0\phantom{0} & 1.85 & 0.664\\
  u40WHW02 & \phantom{0}40 & 40.0\phantom{0} & 1.90 & 0.719\\
  u45WHW02 & \phantom{0}45 & 44.9\phantom{0} & 1.96 & 0.655\\
  u50WHW02 & \phantom{0}50 & 49.8\phantom{0} & 1.83 & 0.574\\
  u60WHW02 & \phantom{0}60 & 59.6\phantom{0} & 1.88 & 0.623\\
  u75WHW02 & \phantom{0}75 & 74.1\phantom{0} & 2.03 & 1.146\\
  \hline
  z20WHW02 & \phantom{0}20 & 20.0\phantom{0} & 1.48 & 0.163\\
  z25WHW02 & \phantom{0}25 & 25.0\phantom{0} & 1.81 & 0.404\\
  z30WHW02 & \phantom{0}30 & 30.0\phantom{0} & 1.50 & 0.221\\
  z35WHW02 & \phantom{0}35 & 35.0\phantom{0} & 1.79 & 0.560\\
  z40WHW02 & \phantom{0}40 & 40.0\phantom{0} & 1.90 & 0.720\\
  \hline
  s25LC06A & \phantom{0}25 & 16.2\phantom{0} & 1.43 & 0.204\\
  s30LC06A & \phantom{0}30 & 12.8\phantom{0} & 1.48 & 0.274\\
  s35LC06A & \phantom{0}35 & 11.8\phantom{0} & 1.48 & 0.242\\
  s40LC06A & \phantom{0}40 & 12.4\phantom{0} & 1.50 & 0.339\\
  s60LC06A & \phantom{0}60 & 16.9\phantom{0} & 1.63 & 0.603\\
  s80LC06A & \phantom{0}80 & 22.4\phantom{0} &  1.67 & 0.628\\
  s120LC06A & 120 & 30.5\phantom{0} & 1.91 & 0.905\\
  \hline
  s40LC06B & \phantom{0}40 & \phantom{0}6.82 & 1.51 & 0.322\\
  s60LC06B & \phantom{0}60 & \phantom{0}5.95 & 1.35 & 0.163\\
  s80LC06B & \phantom{0}80 & \phantom{0}6.04 & 1.46 & 0.185\\
  s120LC06B & 120 & \phantom{0}6.12 & 1.24 & 0.143\\
  \hline 
  s20WH07 & \phantom{0}20 & 15.8\phantom{0} & 1.55 & 0.288\\
  s25WH07 & \phantom{0}25 & 15.8\phantom{0} & 1.60 & 0.334\\
  s30WH07 & \phantom{0}30 & 13.8\phantom{0} & 1.49 & 0.219\\
  s35WH07 & \phantom{0}35 & 13.6\phantom{0} & 1.61 & 0.369\\
  s40WH07 & \phantom{0}40 & 15.3\phantom{0} & 1.83 & 0.599\\
  s45WH07 & \phantom{0}45 & 13.0\phantom{0} & 1.79 & 0.556\\
  s50WH07 & \phantom{0}50 & \phantom{0}9.76 & 1.50 & 0.221\\
  s60WH07 & \phantom{0}60 & \phantom{0}7.25 & 1.46 & 0.175\\
  s80WH07 & \phantom{0}80 & \phantom{0}6.33 &  1.48 & 0.210\\
  s100WH07 & 100 &\phantom{0}6.04 & 1.46 & 0.247 \\
  s120WH07 & 120 & \phantom{0}5.96 & 1.43 & 0.172\\
  \hline
  m35OCWH06 & \phantom{0}35 & 28.1\phantom{0} & 2.08 & 0.457\\
  E20HLW00 & \phantom{0}20 &11.0\phantom{0} & 1.74 & 0.320\\
  E25HLW00 & \phantom{0}25 & \phantom{0}5.45 &1.70 & 0.294
\enddata
\tablecomments{The model name contains the information necessary to
  uniquely specify the presupernova model.  For nonrotating
  progenitors, the beginning letter in the model name refers to the
  metallicity of the progenitor, following the convention of
  \cite{whw:02}, ``$s$'', ``$u$'', and ``$z$'' are used for solar, $10^{-4}$
  solar, and zero metallicities, respectively. Following is the ZAMS
  mass, next we specify the progenitor model set (see the text for
  references). For rotating progenitors, we follow the naming
  convention of the original reference.}  \tablenotetext{a}{We define
  the iron core edge to be where $Y_e$ = 0.495.}
\tablenotetext{b}{$\xi_{2.5}$ is determined at bounce in collapse runs
  using the LS180 EOS and will vary only slightly with EOS.}
  \label{tab:initialmodels}
\end{deluxetable}

One of the most uncertain, yet most important, variables in the
evolution of massive stars is the mass-loss rate.  Mass loss can vary
significantly over the life of a star. Current estimates of mass loss,
either theoretical or based on fits to observational data, can depend
on many parameters, including mass, radius, stellar luminosity,
effective surface temperature, surface hydrogen and helium abundance,
and stellar metallicity \citep{deJager:88,nieuwenhuijzen:90, wellstein:99,
  nugis:00, vink:05}.  The mass-loss rate is uncertain in both the
massive O-star and in the stripped-envelope Wolf-Rayet (W-R) star
stage. O-star winds are expected to be responsible for the partial or
complete removal of the hydrogen envelopes of massive stars. Recent
observational results suggest that the rates used in current stellar
evolution models may be too high by factors of $3-10$ if clumped winds
are considered correctly \citep{bouret:05, fullerton:06,
  puls:06}. With the reduced rates, WR stars would be difficult to
make in standard single-star evolution and would require binary or
eruptive mass-loss scenarios \citep{smith:10}.

\begin{figure}[t]
\begin{center}
\includegraphics[width=\columnwidth]{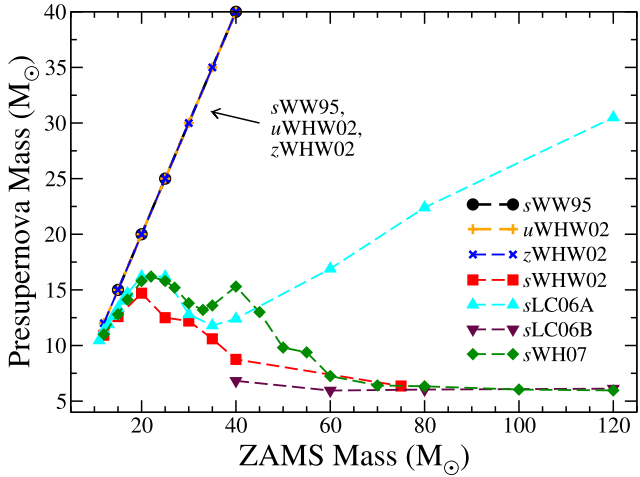} 
\caption{Presupernova mass as a function of ZAMS mass for the various
  model sets considered here. See the text for discussion.
}\label{fig:presupernovamass}
\end{center}
\end{figure}

In Figure~\ref{fig:presupernovamass}, we plot the mass-loss-induced
mapping between ZAMS mass and presupernova mass for the ensemble of
nonrotating progenitors listed in Table~\ref{tab:initialmodels}. WW95
models do not include mass loss - the presupernova models of this
study have a mass equal to the ZAMS mass.  WHW02 and WH07 employ the
mass-loss rates of \cite{nieuwenhuijzen:90} and \cite{wellstein:99}
and use significantly reduced rates for low and zero metallicity
stars.  The $u$ and $z$ models of WHW02 have almost no mass loss and
their presupernova masses are very close to their ZAMS values.  The
solar-metallicity stars of the $s$WHW02 and $s$WH07 model sets have
significant mass loss, generally scaling with ZAMS mass. The most
massive stars in these model sets have presupernova masses that are a
small fraction of the initial ZAMS mass.  For main sequence and giant
phases, \cite{limongi:06} adopt mass-loss rates following
\cite{vink:00,vink:01} and \cite{deJager:88}. For W-R stars, they
either use the mass-loss rates of \cite{nugis:00} (hereinafter
referred to as LC06A models) or \cite{langer:89} (LC06B models).  The
latter are close to the values used for solar-metallicity stars in the
WHW02 and WH07 model sets.  The difference in the LC06A and LC06B
mass-loss rates is roughly a factor of two. This, as portrayed by
Figure~\ref{fig:presupernovamass} and evident from
Table~\ref{tab:initialmodels}, can significantly alter the total mass
at the onset of collapse and also has a strong effect on the iron core
mass and bounce compactness.

An additional uncertainty in massive star evolution is the phenomenon
of large episodic mass loss \citep{smith:08}.  Unknowns and
uncertainties in both the cause and effect of large episodic mass loss
currently prevent detailed stellar evolution calculations from
including this phenomenon at all. 

\subsection{Grid Setup}

Based on resolution studies, we employ a computational grid setup with
a total of 1050 zones.  Near the origin and extending out to $20\,$km,
we employ a constant grid spacing of $80\,$m (250 zones).  Outside of
$20\,$km we logarithmically space the remaining 800 zones to a radius
where the initial density falls to $2000\,$g$\,$cm$^{-3}$.  We require
the high resolution near the center for late postbounce times when the
postshock region becomes small ($r_\mathrm{shock} \lesssim$ $20\,$km)
and when the PNS is close to dynamical collapse to a BH.  We
interpolate the various presupernova profiles ($\rho$, $T$, $Y_e$,
$v$, $\Omega$) to our grid using linear interpolation.

\subsection{Rotation}
In simulations including 1.5D rotation, we directly use
the angular velocity of the progenitor model if it was evolved
with rotation or assign specific angular momentum via the
rotation law
\begin{equation}
  j (r) = j_{16,\infty}  \left[ 1 + \left({A_{M_\odot} \over
        r}\right)^2\right]^{-1} 10^{16}\ \mathrm{cm}^2\ \mathrm{s}^{-1}\,\,,
 \label{eq:specificj}
\end{equation}
where $j_{16,\infty}$ is the specific angular momentum at infinity in
units of $10^{16}\ \mathrm{cm}^2\ \mathrm{s}^{-1}$.  We define
$A_{M_\odot}$ to be the radius where the enclosed mass is
$1\,M_\odot$.  This is a variation on the rotation law commonly used
in simulations of rotating core collapse (e.g., \citealt{ott:06spin}),
where $\Omega(r) = j(r) r^{-2}$ is prescribed and the
differential-rotation parameter $A$ is set to some constant radius.
The advantage of prescribing $j$ (which is conserved along Lagrangian
trajectories) and choosing the value of $A$ based on a mass coordinate
is that progenitors from different groups that are evolved to
different points still yield similar PNS angular momentum
distributions for a given choice of $j_{16,\infty}$.
Equation~(\ref{eq:specificj}) leads to roughly uniform rotation in the
core inside $A_{M_\odot}$ ($j(r)\propto r^2$) and angular velocity
$\Omega(r)$ decreasing with $r^2$ further out ($j(r) =
\mathrm{const}$).  We note that when $1\,M_\odot$ of material is
contained within $10^3\,$km, which is typical of many progenitors, the
central rotation rate is $j_{16,\infty}\,$rad$\,$s$^{-1}$.

Our way of assigning rotation to precollapse models
approximates well the predictions of core rotation (inner $\sim$few
$M_\odot$) from stellar evolution studies (see, e.g.,
\citealt{ott:06spin} for comparison plots) and, thus, is useful for
studying rotational effects on BH formation. Equation~(\ref{eq:specificj})
does not, however, capture the rise in specific angular momentum
observed at larger radii (or mass coordinate) that is important for
the potential evolution toward a long GRB and seen in recent
rotating progenitor models (e.g., \citealt{woosley:06}).

\section{Results}
\label{sec:results}

\subsection{Fiducial Model}
\label{sec:fid}

\begin{figure}[t]
\begin{center}
 \includegraphics[width=\columnwidth]{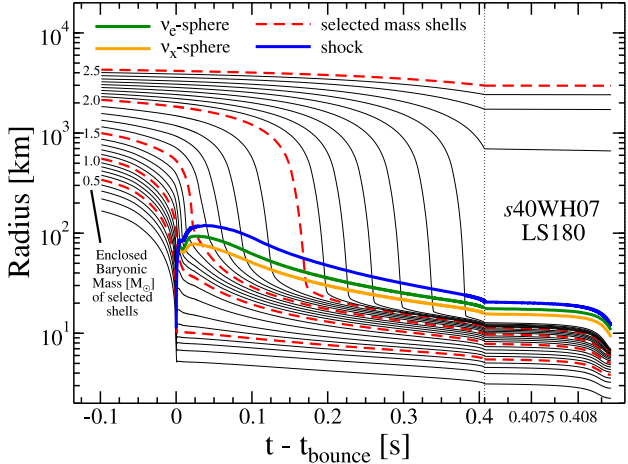}
 \caption{Evolution of baryonic mass shells in the nonrotating model
    $s$40WH07 evolved with the LS180 EOS. We also include the shock
    location and the radii of the $\nu_e$ and $\nu_x$ neutrinospheres.
    The $\bar{\nu}_e$-sphere (not shown), is inside, but very close to
    the $\nu_e$-sphere.  The vertical dotted line denotes a change of
    timescale in the plot, highlighting the final $\sim 1\,$ms of
    evolution before the central density reaches $\sim 4.2
    \times10^{15}\,$g$\,$cm$^{-3}$ and the simulation halts.  We
    specifically highlight the 0.5, 1.0, 1.5, 2.0, and $2.5\,M_\odot$
    baryonic mass shells with dashed lines. With solid lines, for $M <
    2\,M_\odot$, we plot every 0.1 $M_\odot$ mass shell.  Above
    $2\,M_\odot$, we plot mass shells with a spacing of
    $0.05\,M_\odot$.}\label{fig:massshells}
\end{center}
\end{figure}

We begin our discussion with a detailed description of the evolution
of a failing CCSN from core collapse, through bounce, and the
subsequent postbounce evolution to BH formation.  For this, we choose
the 40 $M_\odot$ ZAMS-mass progenitor model $s$40WH07.  We evolve this
progenitor using the LS180 EOS, do not include rotation, and use the
standard setting of $f_\mathrm{heat} = 1$ (see
Section \ref{sec:neutrino}). In Figure~\ref{fig:massshells}, we show the
evolution of the radial coordinate of select baryonic mass shells as
a function of time and we highlight shells enclosing 0.5, 1.0, 1.5,
2.0, and $2.5\,M_\odot$.  In addition, the figure shows the shock
radius and the positions of the energy-averaged $\nu_e$ and $\nu_x$
neutrinospheres as a function of time.  The prebounce collapse phase
($t < 0$) lasts $\sim 450\,$ms. At bounce, the central value of the
lapse function is $\alpha_c \sim 0.82$, and the metric function $X$
has a maximum of $\sim 1.1$ and peaks off-center at a baryonic mass
coordinate of $\sim 0.6\,M_\odot$ which roughly corresponds to the
edge of the inner core.  The inner core initially overshoots to a
maximum central density $\rho_c \sim
5.0\times10^{14}\,$g$\,$cm$^{-3}$, then settles at $\sim 3.7 \times 10
^{14}\,$g$\,$cm$^{-3}$. $\rho_c$ subsequently increases as accretion
adds mass to the PNS. The bounce shock forms at a baryonic mass
coordinate of $\sim 0.6\,M_\odot$.  From there, it moves out quickly
in mass, reaching a baryonic mass coordinate of $\sim 1.5 \, M_\odot$
at $22\,$ms after bounce, $2\,M_\odot$ at $\sim 162\,$ms, and
$2.25\,M_\odot$ at $\sim 329\,$ms. In radius, the shock reaches a
maximum of $\sim 120\,$km at $38\,$ms after bounce.  There it stalls,
then slowly recedes. At $10\,$ms after bounce, the accretion rate
through the shock is $\sim 18\,M_\odot\,$s$^{-1}$ and drops to $\sim
2.7$, $\sim 1.7$, and $\sim 1.25\,M_\odot\,$s$^{-1}$ at 100, 200, and
$300\,$ms after bounce, respectively.  The drop in the accretion rate
has little effect on the failing supernova engine. In agreement with
previous work that employed a more accurate neutrino treatment
(e.g., \citealt{thompson:03, liebendoerfer:05}), the 1D neutrino
mechanism is manifestly ineffective in driving the shock, yielding, in
this model, a heating efficiency $\eta = L_\mathrm{absorbed} /
L_{\nu_e + \bar{\nu}_e}$ of only $\sim 3\,$\% (on average).  The
neutrinospheres (where the energy-averaged optical depth $\tau = 2/3$)
are initially exterior to the shock but are surpassed by the latter in
a matter of milliseconds after bounce, leading to the $\nu_e$
deleptonization burst.  At all times, the $\nu_x$ neutrinosphere is
interior to the $\bar{\nu}_e$ neutrinosphere, which in turn is
slightly interior to the $\nu_e$ neutrinosphere.  The mean neutrino
energies also follow this order.  They are the largest for $\nu_x$ and
the lowest for $\nu_e$ and increase with decreasing neutrinosphere
radii (e.g., \citealt{thompson:03,sumiyoshi:07,ott:08,fischer:09a}).

At $\sim 408\,$ms after bounce, the shock has receded to $\sim 20\,$km
and the PNS has reached a baryonic (gravitational) mass of $\sim
2.33\,M_\odot$ ($\sim 2.23\,M_\odot$). The difference between baryonic
and gravitational mass, at this point in the evolution, is due to the
$\sim 1.9 \times 10^{53}\,$erg of energy radiated by neutrinos.  At
this point, dynamical PNS collapse to a BH sets in and happens on a
coordinate timescale of $\lesssim 1\,$ms.  In the rightmost part of
Figure~\ref{fig:massshells}, we zoom in to the final $1\,$ms of
evolution to show detail.  The first signs of collapse manifest
themselves in the development of a radial infall velocity profile at
the PNS edge.  The PNS then collapses in on itself and the central
density increases by a factor of $\sim 3$ in only $\sim 1\,$ms of
coordinate time.  The simulation crashes due to EOS limitations at
$\rho_c \sim 4.2\times10^{15}\,$g$\,$cm$^{-3}$ and with $\alpha_c =
0.006$. At this point the peak of the metric function $X =
[1-2m(r)/r]^{-1/2}$ is $\sim 4.4$ at a coordinate radius of $\sim
6.8\,$km.  There, the fluid velocity also peaks at $\sim -\,0.83\,c$.
The shock recedes by $\sim 8\,$km in the last $\sim 1\,$ms of
evolution to a radial coordinate of $\sim 12\,$km.  During the last
$\sim 0.05\,$ms, due to the central lapse dropping to nearly zero, the
evolution of the mass shells slows near the origin.  This is
characteristic for our choice of gauge. If the simulation were to
continue, $X$ would become singular at the event horizon that would
appear after infinite coordinate time in our coordinates
\citep{petrich:86}.

The $s$40WH07 model discussed here is a typical example of a failing
CCSN in spherical symmetry. We present the results of a large number
of such models in Table~\ref{tab:results}, where for each EOS and
progenitor model we show the time to BH formation as measured from
bounce and the mass, both baryonic and gravitational, of the PNS when
the central value of the lapse function $\alpha$ reaches 0.3 (roughly
the point of instability). In this table, the model name describes the
initial model.  The metallicity is denoted by one of three letters:
$s$, $u$, and $z$ which represent solar, $10^{-4}$ solar, and zero
metallicity, respectively.  Following the metallicity is the ZAMS mass
and the progenitor model set.  In many simulations, particularly in
those employing stiff EOS, a BH does not form within $3.5\,$s.  For
these simulations we include in parentheses the mass inside the shock
at $3.5\,$s. We note that at BH formation the shock is typically at a
distance of $\lesssim 20\,$km and there is very little mass between
the shock and the PNS. The dynamical collapse to a BH happens very
quickly ($t \lesssim 1\,$ms) during which very little additional
accretion occurs.

\begin{deluxetable*}{rccccccccccccccc}
\tablecolumns{17}
\tablewidth{0pc}
\tablecaption{Black Hole Formation Properties}
\tablehead{
Model & \multicolumn{3}{c}{LS180} && \multicolumn{3}{c}{LS220}
&&\multicolumn{3}{c}{LS375}  && \multicolumn{3}{c}{HShen}  \\
&$t_\mathrm{BH} $ & $M_\mathrm{b,max} $& $M_\mathrm{g,max} $ &&$t_\mathrm{BH}$ & $M_\mathrm{b,max}$&
$M_\mathrm{g,max}$ &&$t_\mathrm{BH} $ & $M_\mathrm{b,max} $& $M_\mathrm{g,max} $ &&$t_\mathrm{BH}$ & $M_\mathrm{b,max} $& $M_\mathrm{g,max} $ \\
& (s) & ($M_\odot$) & ($M_\odot$) && (s) & ($M_\odot$) & ($M_\odot$) && (s) & ($M_\odot$) & ($M_\odot$) &&(s) & ($M_\odot$) & ($M_\odot$) }
\startdata
$s$20WW95   &  0.787 &  2.238 &  2.108 &&  1.129 &  2.377 &  2.201 && 3.351 & 3.060 & 2.653 && 2.287 &  2.751 &  2.486 \\
$s$25WW95   &  0.737 &  2.246 &  2.118 &&  1.046 &  2.383 &  2.211 && 2.707 & 3.054 & 2.656 && 1.990 &  2.760 &  2.498 \\
$s$40WW95   &  0.524 &  2.263 &  2.137 & & 0.666 &  2.406 &  2.240 && 1.381 & 3.043 & 2.674 && 1.129 &  2.815 &  2.562 \\
\hline
\hline
$s$20WHW02  &  $\cdots$ &  (1.949) &  (1.794) &&  $\cdots$ &  (1.950) &  (1.798) &&  $\cdots$ &  (1.951) &  (1.807) &&  $\cdots$ &  (1.943) &  (1.805) \\
$s$25WHW02  &  1.021 &  2.211 &  2.079 &&  1.504 &  2.355 &  2.172 && $\cdots$ &  (2.917) &  (2.559) &&  2.929 &  2.736 &  2.468 \\
$s$30WHW02  &  1.820 &  2.144 &  1.978 &&  2.986 &  2.331 &  2.108 &&  $\cdots$ &  (2.416) &  (2.182) &&  $\cdots$ &  (2.405) &  (2.190) \\
$s$35WHW02  &  2.073 &  2.141 &  1.976 &&  3.334 &  2.328 &  2.105 &&  $\cdots$ &  (2.351) &  (2.137) &&  $\cdots$ &  (2.340) &  (2.141) \\
$s$40WHW02  &  1.512 &  2.168 &  2.019 &&  2.231 &  2.336 &  2.134 &&  $\cdots$ &  (2.634) &  (2.355) &&  $\cdots$ &  (2.615) &  (2.362) \\
$s$75WHW02  &  $\cdots$ &  (1.920) &  (1.781) &&  $\cdots$ &  (1.920) &  (1.784) && $\cdots$ &  (1.921) &  (1.791) &&  $\cdots$ &  (1.913) &  (1.787) \\
\hline
$u$20WHW02  &  0.938 &  2.215 &  2.082 &&  1.367 &  2.358 &  2.175 && $\cdots$ &  (2.852) &  (2.516) &&  3.004 &  2.734 &  2.466 \\
$u$25WHW02  &  1.759 &  2.160 &  2.009 &&  2.798 &  2.330 &  2.124 && $\cdots$ &  (2.446) &  (2.218) && $\cdots$ &  (2.429) &  (2.220) \\
$u$30WHW02  &  0.922 &  2.217 &  2.084 &&  1.353 &  2.359 &  2.178 &&  $\cdots$ &  (2.802) &  (2.483) &&  3.228 &  2.731 &  2.462 \\
$u$35WHW02  &  0.379 &  2.347 &  2.242 &&  0.484 &  2.465 &  2.329 &&  1.308 &  3.020 &  2.693 &&  1.075 &  2.847 &  2.608 \\
$u$40WHW02  &  0.369 &  2.346 &  2.241 &&  0.453 &  2.469 &  2.333 &&  0.946 &  3.023 &  2.710 &&  0.849 &  2.874 &  2.638 \\
$u$45WHW02  &  0.441 &  2.301 &  2.187 &&  0.548 &  2.433 &  2.284 &&  1.108 &  3.027 &  2.694 &&  0.959 &  2.842 &  2.600 \\
$u$50WHW02  &  0.563 &  2.273 &  2.154 &&  0.706 &  2.408 &  2.251 &&  1.365 &  3.030 &  2.676 &&  1.163 &  2.816 &  2.569 \\
$u$60WHW02  &  0.432 &  2.363 &  2.267 &&  0.579 &  2.460 &  2.331 &&  1.346 &  3.009 &  2.693 &&  1.165 &  2.849 &  2.620 \\
$u$75WHW02  &  0.226 &  2.526 &  2.449 &&  0.285 &  2.592 &  2.498 &&  0.626 &  3.006 &  2.775 &&  0.594 &  2.984 &  2.791 \\
\hline
$z$20WHW02  &  3.295 &  2.116 &  1.934 &&  $\cdots$ &  (2.141) &  (1.955) && $\cdots$ &  (2.143) &  (1.968) &&  $\cdots$ &  (2.132) &  (1.968) \\
$z$25WHW02  &  0.602 &  2.283 &  2.167 &&  0.956 &  2.398 &  2.239 & & 3.443 &  3.050 &  2.650 & &  2.351 &  2.762 &  2.505 \\
$z$30WHW02  &  1.772 &  2.149 &  1.989 &&  2.964 &  2.329 &  2.114 &&  $\cdots$ &  (2.413) &  (2.187) &&  $\cdots$ &  (2.401) &  (2.192) \\
$z$35WHW02  &  0.446 &  2.321 &  2.213 &&  0.619 &  2.436 &  2.291 &&  1.939 &  3.027 &  2.669 &&  1.380 &  2.813 &  2.569 \\
$z$40WHW02  &  0.365 &  2.350 &  2.245 &&  0.450 &  2.471 &  2.335 &&  0.958 &  3.023 &  2.711 & & 0.856 &  2.874 &  2.639 \\
\hline
\hline
$s$25LC06A  &  1.220 &  2.176 &  2.029 &&  2.547 &  2.333 &  2.130 &&  $\cdots$ &  (2.440) &  (2.213) &&  $\cdots$ &  (2.398) &  (2.195) \\
$s$30LC06A  &  1.101 &  2.181 &  2.035 && 1.726 &  2.342 &  2.141 &&  $\cdots$ &  (2.767) &  (2.446) &&  $\cdots$ &  (2.695) &  (2.421) \\
$s$35LC06A  &  1.029 &  2.186 &  2.040 &&  1.726 &  2.338 &  2.133 &&  $\cdots$ &  (2.567) &  (2.305) &&  $\cdots$ &  (2.517) &  (2.285) \\
$s$40LC06A  &  0.746 &  2.232 &  2.102 &&  1.138 &  2.372 &  2.193 && $\cdots$ &  (2.796) &  (2.470) &&  3.390 &  2.723 &  2.452 \\
$s$60LC06A  &  0.393 &  2.331 &  2.224 &&  0.512 &  2.450 &  2.310 &&  1.536 &  3.025 &  2.678 &&  1.278 &  2.816 &  2.572 \\
$s$80LC06A  &  0.429 &  2.308 &  2.197 &&  0.530 &  2.437 &  2.293 &&  1.075 &  3.021 &  2.689 &&  1.083 &  2.825 &  2.581 \\
$s$120LC06A &  0.262 &  2.439 &  2.351 &&  0.317 &  2.531 &  2.423 &&  0.661 &  3.001 &  2.745 &&  0.728 &  2.911 &  2.701 \\
\hline
$s$40LC06B  &  0.958 &  2.189 &  2.043 &&  1.411 &  2.349 &  2.152 && $\cdots$ &  (2.957) &  (2.576) &&  2.887 &  2.720 &  2.444 \\
$s$60LC06B  &  3.073 &  2.117 &  1.934 &&  $\cdots$ &  (2.166) &  (1.972) &&  $\cdots$ &  (2.165) &  (1.984) &&  $\cdots$ &  (2.126) &  (1.961) \\
$s$80LC06B  &  2.441 &  2.131 &  1.963 &&  $\cdots$ &  (2.260) &  (2.052) &&  $\cdots$ &  (2.264) &  (2.071) &&  $\cdots$ &  (2.249) &  (2.069) \\
$s$120LC06B &  2.983 &  2.120 &  1.944 &&  $\cdots$ &  (2.171) &  (1.984) &&  $\cdots$ &  (2.167) &  (1.992) &&  $\cdots$ &  (2.102) &  (1.947) \\
\hline
\hline
$s$20WH07   &  1.275 &  2.180 &  2.035 &&  1.876 &  2.341 &  2.143 && $\cdots$ &  (2.712) &  (2.412) &&  $\cdots$ &  (2.694) &  (2.426) \\
$s$25WH07   &  1.066 &  2.202 &  2.065 &&  1.523 &  2.352 &  2.165 && $\cdots$ &  (2.975) &  (2.595) &&  2.796 &  2.736 &  2.466 \\
$s$30WH07   &  1.751 &  2.150 &  1.991 &&  2.978 &  2.329 &  2.115 &&  $\cdots$ &  (2.408) &  (2.184) &&  $\cdots$ &  (2.397) &  (2.190) \\
$s$35WH07   &  0.836 &  2.232 &  2.104 &&  1.203 &  2.369 &  2.194 && $\cdots$ &  (2.918) &  (2.563) &&  2.689 &  2.744 &  2.481 \\
$s$40WH07   &  0.408 &  2.334 &  2.228 &&  0.561 &  2.448 &  2.306 &&  1.596 &  3.024 &  2.680 &&  1.259 &  2.827 &  2.585 \\
$s$45WH07   &  0.454 &  2.319 &  2.210 &&  0.626 &  2.435 &  2.289 &&  2.027 &  3.028 &  2.667 &&  1.395 &  2.812 &  2.567 \\
$s$50WH07   &  1.813 &  2.147 &  1.987 &&  2.989 &  2.329 &  2.113 &&  $\cdots$ &  (2.411) &  (2.185) &&  $\cdots$ &  (2.399) &  (2.190) \\
$s$60WH07   &  2.778 &  2.124 &  1.947 &&  $\cdots$ &  (2.230) &  (2.023) &&  $\cdots$ &  (2.232) &  (2.039) &&  $\cdots$ &  (2.220) &  (2.040) \\
$s$80WH07   &  2.113 &  2.139 &  1.974 &&  3.284 &  2.328 &  2.104 &&  $\cdots$ &  (2.363) &  (2.145) &&  $\cdots$ &  (2.350) &  (2.148) \\
$s$100WH07  &  1.457 &  2.163 &  2.008 &&  2.355 &  2.335 &  2.124 && $\cdots$ &  (2.539) &  (2.281) &&  $\cdots$ &  (2.524) &  (2.289) \\
$s$120WH07  &  3.043 &  2.120 &  1.940 &&  $\cdots$ &  (2.179) &  (1.985) &&  $\cdots$ &  (2.180) &  (1.999) &&  $\cdots$ &  (2.169) &  (1.999)
\enddata
\tablecomments{BH formation times and maximum PNS mass (both baryonic
  and gravitational) for nonrotating runs with $f_{\mathrm{heat}} = 1$
  for all four EOS. We stop our simulations at $3.5\,$s after core
  bounce.  Models that have not formed a BH by then probably explode
  in nature.  They are marked by $\cdots$, but we include the masses
  inside the shock at $3.5\,$s in parentheses. The progenitor models
  are the result of various stellar evolution studies: WW95,
  \cite{ww:95}; WHW02, \cite{whw:02}; LC06, \cite{limongi:06}; and
  WH07, \cite{woosley:07}. }
\label{tab:results}
\end{deluxetable*}

\begin{deluxetable}{rllll}
  \tablecolumns{5} 
  \tablewidth{0pc}
  \tablecaption{$s$40WW95 Comparison with LS180 and HShen EOS.}
  \tablehead{\multirow{3}{*}{Study} & \multicolumn{2}{c}{LS180} & \multicolumn{2}{c}{HShen} \\
    & $t_\mathrm{BH}$ & $M_{b,\mathrm{max}}$ & $t_\mathrm{BH}$ & $M_{b,\mathrm{max}}$ \\
    & ($s$) & ($M_\odot$) &($s$) & ($M_\odot$) } 
\startdata
\cite{liebendoerfer:04} & $\sim$0.5\phantom{3} &$\sim$2.20\phantom{986} & $\cdots$ & $\cdots$ \\
\cite{sumiyoshi:07} & \phantom{$\sim$}0.56\phantom{55}& \phantom{$\sim$}2.1\phantom{96} & 1.34 & 2.66 \\
\cite{fischer:09a} & \phantom{$\sim$}0.4355& \phantom{$\sim$}2.196 & 1.030\tablenotemark{a} & 2.866\tablenotemark{a}\\
This work & \phantom{$\sim$}0.524\phantom{5}& \phantom{$\sim$}2.263 & 1.129 & 2.815
\enddata
\tablenotetext{a}{See the text for a discussion of the HShen EOS results
from \cite{fischer:09a}.}
\label{tab:compare_s40}
\end{deluxetable}

\subsection{Comparison with Previous Work}
\label{sec:previousworks}

The $s$40WW95 progenitor was considered in the BH formation studies of
\cite{liebendoerfer:04}, \cite{sumiyoshi:07} (hereinafter referred to
as S07), and \cite{fischer:09a} (hereinafter referred to as F09). For
comparison, we perform simulations with this progenitor for both the
LS180 and HShen EOS.  Table~\ref{tab:compare_s40} compares two key
quantities, the time to BH formation and the maximum baryonic PNS
mass, obtained with \code{GR1D} with the results obtained in the
aforementioned studies.
 
For the LS180 EOS, we find a time to BH formation of $\sim 524\,$ms
and a maximum baryonic PNS mass of $\sim 2.26\,M_\odot$, which is
$\sim 3\,$\% larger than predicted by F09.  We attribute this
discrepancy to the different neutrino transport methods
used. \code{GR1D}'s leakage scheme has the tendency to somewhat
over predict electron-type neutrino luminosities (see the discussion
in \citealt{oconnor:10}), resulting in lower gravitational masses
compared to full Boltzmann transport calculations. Our time to BH
formation is longer by $\sim 100\,$ms or $\sim 20\,$\%.  This
disagreement is relatively larger than the baryonic mass disagreement
due to the low accretion rate at late times that translates small
differences in mass to large differences in time. At $\sim 435.5\,$ms,
the time to BH formation of F09, our PNS has a baryonic mass of $\sim
2.17\,M_\odot$, which is consistent to $\sim 1\,$\% with F09.  We find
it more difficult to reconcile our results (and those of
\cite{liebendoerfer:04} and F09) with the simulations of S07.  Their
maximum PNS baryonic mass and the time to BH formation suggest a lower
accretion rate throughout their evolution ($\sim 2.1 \, M_\odot$ in
$\sim 560\,$ms).

In the simulation run with the stiffer HShen EOS, the larger maximum
PNS mass leads to a delay of BH formation until a postbounce time
$\sim1.129\,$s and we find a maximum baryonic PNS mass of $\sim
2.82\,M_\odot$.  The maximum PNS mass and time to BH formation of S07
again suggest an accretion rate in disagreement with F09 and our work.
The results of F09 with the HShen EOS suffer from a glitch in F09's
EOS table interpolation scheme which has since been fixed (T. Fischer
2010, private communication). This leads to a postbounce time to BH
formation of $\sim 1.4\,$s and a maximum baryonic PNS mass of $\sim
3.2 \, M_\odot$. Results from more recent simulations correct this
error and are presented in Table~\ref{tab:compare_s40} (T. Fischer
2010, private communication).

\subsection{Equation-of-state Dependence and Thermal Effects}
\label{sec:eosdependence}

\begin{figure}[t]
\begin{center}
 \includegraphics[width=\columnwidth]{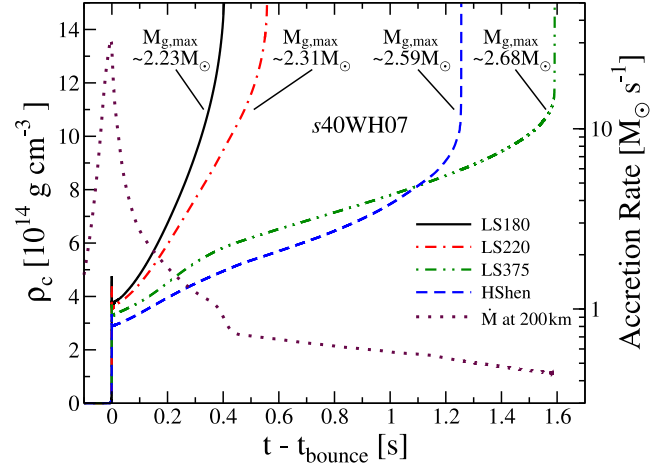}
 \caption{Central density (left ordinate) and accretion rate (right
    ordinate) vs. time since bounce for the $s$40WH07 progenitor and
    four EOS.  BH formation occurs when the central density diverges.
    Each $\rho_c$ curve is annotated with the maximum gravitational
    PNS mass. The drop in the accretion rate at $t \sim 0.4\,$s is due
    to the accretion of a mass shell where the density drops by $\sim
    30\,$\%. Note the accretion rate is on a logarithmic
    scale.}\label{fig:s40WH07_rhoc}
\end{center}
\end{figure}

The maximum PNS mass and, thus, the evolution toward BH formation,
depends strongly on the EOS.  This was realized early on
\citep{burrows:88} and has recently been investigated by S07 and F09
who compared models evolved with the LS180 and HShen EOS.  Here we
extend their discussion and include also the LS220 and LS375 EOS.
For a given accretion history, set by progenitor structure and
independent of the high-density EOS, a stiffer nuclear EOS leads to
a larger postbounce time to BH formation. In
Figure~\ref{fig:s40WH07_rhoc}, we plot the evolution of the central
density $\rho_c$ of the $s$40WH07 model evolved with the four
considered EOS.  Each EOS leads to a characteristic maximum central
density at bounce that is practically independent of progenitor
model: $\sim 4.8 \times 10^{14}\,$g$\,$cm$^{-3}$, $\sim 4.4 \times
10^{14}\,$g$\,$cm$^{-3}$, $\sim 3.7 \times 10^{14}\,$g$\,$cm$^{-3}$,
and $\sim 3.4 \times 10^{14}\,$g$\,$cm$^{-3}$ for the LS180, LS220,
LS375, and HShen EOS, respectively. As expected, the variant using
the softest nuclear EOS (LS180) shows the steepest postbounce
increase in $\rho_c$ and becomes unstable to BH formation at only
$\sim 408\,$ms for this progenitor. The onset of BH formation is
marked by a quick rise in the central density. This is most obvious
from the $\rho_c$ evolutions of the model variants using the stiff
HShen and LS375 EOS.

Interestingly, the nominally stiffest EOS (LS375) leads to higher
central densities than the softer HShen EOS up until $\sim 1.1\,$s
after bounce.  We find that this is due to the HShen EOS yielding
higher pressure at $\rho \lesssim 3\times10^{14}\,$g$\,$cm$^{-3}$,
$T\sim 10\,$MeV, and $Y_e\sim 0.3$. This higher pressure, initially in
the core and later in the outer PNS layers, maintains the PNS at a
lower central density.  The cold-NS mass-radius relation shown in
Figure~\ref{fig:MvsRtemp0.1} also exhibits this. For a given low-mass
NS, the HShen EOS predicts a lower central density.  For cold NSs,
this trend continues until $\rho_c \sim 5.4 \times
10^{14}\,$g$\,$cm$^{-3}$. Thermal effects, which are stronger in the
HShen EOS, will increase this value for hot PNSs.

We also plot in Figure~\ref{fig:s40WH07_rhoc} the evolution of the
mass accretion rate $\dot{M}$ in model $s$40WH07 (evaluated at a
radius of $200\,$km). Variations in the high-density EOS have no
effect on $\dot{M}$ which is most sensitive to progenitor
structure. Sudden drops in $\dot{M}$ occur when density
discontinuities that go along with compositional interfaces advect
in. An example of this can be seen in $s$40WH07 at $\sim 400\,$ms
after bounce where $\dot{M}$ drops by $\sim 30\,$\% due to a density
jump at a baryonic mass coordinate of $\sim 2.35\,M_\odot$.  Such
interfaces are common features of evolved massive stars \citep{whw:02}
and can help jumpstart shock revival in special cases (see, e.g., the
11.2 $M_\odot$ model of \cite{buras:06b}, and Section \ref{sec:lums} of this
study).  In the BH-formation context, they lead to a disproportionate
increase in the time to BH formation in models whose EOS permit a PNS
with mass greater than the mass coordinate of the density jump.

The maximum gravitational (baryonic) PNS masses for the four models
shown in Figure~\ref{fig:s40WH07_rhoc} are $\sim 2.23\,M_\odot$ ($\sim
2.33\,M_\odot$), $\sim 2.31\,M_\odot$ ($\sim 2.45\,M_\odot$), $\sim
2.68\,M_\odot$ ($\sim 3.02\,M_\odot$), and $\sim 2.59\,M_\odot$ ($\sim
2.83\,M_\odot$); and the BH formation times are $\sim 408\,$ms, $\sim
561\,$ms, $\sim 1.596\,$s, and $\sim 1.259\,$s for the LS180, LS220,
LS375, and HShen EOS, respectively. The maximum cold NS gravitational
masses are $\sim 1.83\,M_\odot$, $\sim 2.04\,M_\odot$, $\sim
2.72\,M_\odot$, and $\sim 2.24\,M_\odot$ for the LS180, LS220, LS375,
and HShen EOS, respectively.

\begin{figure}[t]
\begin{center}
 \includegraphics[width=\columnwidth]{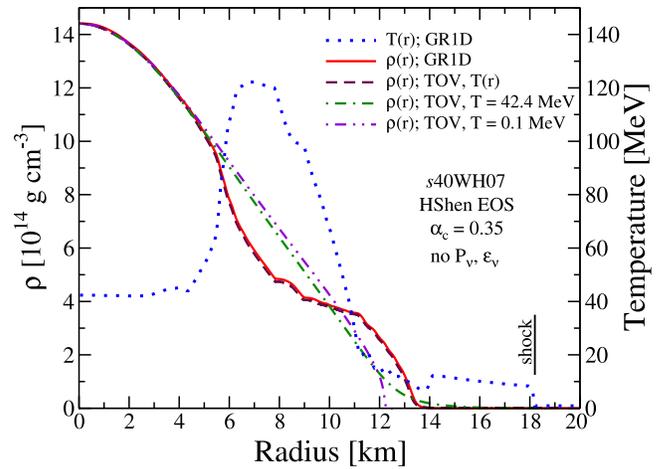}
 \caption{Comparison of radial density (left ordinate) and
    temperature (right ordinate) profiles of the PNS just before
    collapse ($\alpha_c = 0.35$) in model $s$40WH07 evolved with the
    HShen EOS with profiles obtained from a TOV solution using the
    same central density and the same radial temperature and $Y_e$
    distributions as in model $s$40WH07 (dashed). For comparison, we
    also include profiles obtained with the TOV equations assuming
    both $T = T_c = 42.4\,$MeV (dot-dashed) and $T=0.1\,$MeV ``cold''
    (dot-dot-dashed) and $\beta$-equilibrium. The flattening of the
    density profile between 5 and $11\,$km is due to the strong thermal
    pressure support in this region (dotted). The gravitational mass
    inside the shock (whose position we denote with a vertical black
    line) of the $s$40WH07 model and of the TOV star agree to within
    $2\,$\%. For this comparison, we switched off neutrino contributions
    to the internal energy and pressure in
    \code{GR1D}.}\label{fig:TOVvsGR1D}
\end{center}
\end{figure}

In models evolved with the LS180, LS220, and HShen EOS, the maximum
gravitational PNS mass is larger than the maximum gravitational cold
NS mass.  We can understand the differences between these cold NS and
PNS maximum masses by comparing the PNS structure with various TOV
solutions.  In Figure~\ref{fig:TOVvsGR1D}, we plot the density and
temperature profiles of the $s$40WH07 model evolved with the HShen EOS
just prior to collapse to a BH.  At this time, $t \sim 1.098\,$s, the
central lapse is $\alpha_c = 0.35$, the central density is $\rho_c
\sim 1.44\times 10^{15}\,$g$\,$cm$^{-3}$, $T_c \sim 42.4\,$MeV, and
the PNS gravitational (baryonic) mass is $\sim 2.51\,M_\odot$ ($\sim
2.74\,M_\odot$).  For comparison, we include in
Figure~\ref{fig:TOVvsGR1D} three TOV solutions, all with the same
central density but different temperature and $Y_e$ profiles.
Specifically, we plot the density profile assuming (1)
$T(r)=0.1\,$MeV, this is the ``cold'' NS case, (2)
$T(r)=42.4\,$MeV, which corresponds to the central temperature from
the \code{GR1D} evolution, and (3) $T(r) = T_{\code{GR1D}}$,
assuming the same radial temperature profile as the \code{GR1D} model.
We impose neutrinoless $\beta$ equilibrium for the former two TOV
solutions and, similar to the temperature, assume the $Y_e$ profile of
the \code{GR1D} model for the latter. For this comparison, \code{GR1D}
is run without neutrino pressure and energy contributions, since they
are also neglected in the TOV solution.

Inside of $\sim 6\,$km, corresponding to a gravitational mass
coordinate of $\sim 0.4\,M_\odot$, the material is not shock heated
but rather is heated only via adiabatic compression.  The outer
regions ($\sim 6-11\,$km) of the PNS are hot compared to the inner
core. This is due to accretion and compression of shock heated
material onto the PNS surface.  In this region, the thermal pressure
support is sufficiently strong to flatten out the PNS density profile.
More mass is located at larger radii compared to constant-temperature
TOV solutions.  This decreases PNS compactness, increasing the maximum
gravitational mass.  The cold-NS and the $T=T_c$ TOV solutions have a
gravitational mass of $\sim 2.23\,M_\odot$ and $\sim 2.35\,M_\odot$,
respectively. On the other hand, the TOV solution that assumes the
same $T$ and $Y_e$ profile as the \code{GR1D} model yields a
gravitational mass of $\sim 2.46\,M_\odot$, within $\sim 2\,$\% of the
PNS gravitational mass in the full \code{GR1D} simulation. Tests in
which we vary the $Y_e$ distribution in the TOV solution with $T=T(r)$
show that the maximum PNS mass is insensitive to variations in $Y_e$
from the \code{GR1D} profile to neutrinoless $\beta$-equilibrium.  All
this leads us to the conclusion that it is thermal pressure
  support in the outer PNS core that is responsible for increasing the
  maximum stable gravitational PNS mass beyond that of a cold NS.
Our finding is in agreement with the recent BH formation simulations
of \cite{sumiyoshi:07} and \cite{fischer:09a}, who noted the same
differences to cold TOV solutions, but did not pinpoint their precise
cause. However, our result is in disagreement with \cite{burrows:88}
who reported maximum PNS masses within a few percent of a solar mass
off their cold-NS values. This could be related to Burrows's specific
choice of EOS.  As we discuss below, stiff nuclear EOS have a more
limited response to thermal effects. Another resolution to this
disagreement could be the nature of his PNS cooling simulations that
were not hydrodynamic, but rather employed a Henyey relaxation
approach with imposed accretion.

We find the same overall systematics of increased maximum PNS mass due
to thermal pressure support for the entire set of progenitors evolved
with the LS180, LS220, and HShen EOS (variations due to differences in
progenitor structure are discussed in Section \ref{sec:prestruct}). In the
sequence of the LS EOS, the relevance of thermal pressure support
decreases with increasing stiffness. In the case of the perhaps
unphysically stiff LS375 EOS, the effect of high temperatures in the
outer PNS core is reversed: the PNSs in \code{GR1D} simulations
become unstable to collapse at lower maximum masses than their
cold counterparts. This very surprising  observation is understood by
considering that in GR, higher temperatures not only add thermal
pressure support to the PNS, but also increase its mass-energy. This
results in a deeper effective potential well and, thus, is
destabilizing. In the LS180, LS220, and HShen case, the added thermal
pressure support is significant and dominates over the latter effect.
In the very stiff LS375 EOS, the added thermal pressure component is
negligible, and the destabilizing effect dominates.

Finally, we point out quantitative differences in models evolved with
and without neutrino pressure in the dense neutrino-opaque core.  In
the $s$40WH07 model evolved with the HShen EOS, the difference in the
maximum gravitational mass is $\sim 0.08\,M_\odot$ ($\sim 3\,$\%) and
the difference in the time to BH formation is $\sim 160\,$ms ($\sim
14\,$\%). These numbers depend on the employed EOS and progenitor
model. In test calculations with a variety of progenitors and EOS, we
generally find increases of the maximum PNS gravitational mass of
$\lesssim 5\,$\%.

\subsection{Influence of Presupernova Structure}
\label{sec:prestruct}

\begin{figure}[t]
\begin{center}
 \includegraphics[width=\columnwidth]{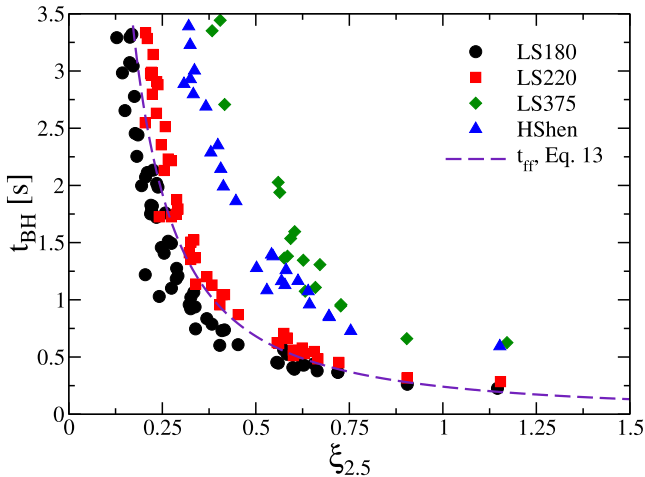} 
 \caption{BH formation time as a function of the bounce compactness
    ($\xi_{2.5}$) for all nonrotating models presented in
    Table~\ref{tab:results} that form BHs within $3.5\,$s of bounce.
    Simulations performed with the LS180, LS220, LS375, and HShen EOS
    are labeled with circles, squares, diamonds, and triangles,
    respectively.  Also shown (dashed line) is the free fall time to
    the origin (Equation~\ref{eq:freefalltime}) of a mass element located
    at a baryonic mass coordinate of
    $2.5\,M_\odot$.}\label{fig:tbh_vs_xi}
\end{center}
\end{figure}

The failure of a CCSN becomes definite only when accretion pushes the
PNS over its maximum mass and a BH forms.  Hence, the time to BH
formation is a hard upper limit to the time available for the
supernova mechanism to reenergize the shock.  We will demonstrate in
this section that it is possible to estimate, for a given nuclear EOS,
the postbounce time to BH formation in non- or slowly spinning on the
basis of a single parameter, progenitor bounce compactness,
$\xi_{2.5}$, which we introduced in Section \ref{sec:initialmodels}.  In
Figure~\ref{fig:tbh_vs_xi}, we plot the postbounce time to BH formation
($t_\mathrm{BH}$) as a function of $\xi_{2.5}$ for all nonrotating
models considered in this study and listed in
Table~\ref{tab:results}. The distribution of data points for each EOS
can be fit with a function $\propto (\xi_{2.5})^{-3/2}$. This
remarkable result can be understood as follows: using Kepler's third
law, consider the Newtonian free fall time to the origin for a mass
element $dm$ initially located at $r_*$ and on a radial orbit about a
point mass of $M^* \gg dm$,
\begin{equation} 
  t_{\mathrm{ff}} = {1\over 2} \sqrt{4\pi^2a^3 \over GM^*} = \pi\sqrt{{r_*^3 \over 8 G M^*}}\,.\label{eq:fftimea}
\end{equation} 

Here, for clarity, the quantities are in cgs units. $G$ is the gravitational
constant and $a$ is the semimajor axis equal to half of the apoapsis,
$r_*$. Recalling the definition of $\xi_{2.5}$, if the mass element
$dm$ is located at a mass coordinate of $2.5\,M_\odot$ and at a radial
coordinate of $r_*$, then $r_* = 2500\,\mathrm{km} / \xi_{2.5}$ , and we can
write the free fall time in terms of $\xi_{2.5}$,
\begin{equation}
 t_{\mathrm{ff}}^{2.5M_\odot} = 0.241(\xi_{2.5})^{-3/2}\,\mathrm{s}.\label{eq:freefalltime}
\end{equation}

In Figure~\ref{fig:tbh_vs_xi}, we overplot this Newtonian free fall time
for a mass element at baryonic mass coordinate $2.5\,M_\odot$ as a
function of $\xi_{2.5}$. For small $\xi_{2.5}$, the mass element
begins its free fall from a large radius and, hence, takes longer to
reach to origin. In general, material in outer layers of the star will
not begin to fall freely until it loses pressure support. Hence, the
free fall approximation is not exact (within a factor of $\sim 2$;
\citealt{burrows:86bh}), but describes the general behavior of
$t_\mathrm{BH}$ very well. The deviation of data points from the free
fall curve is because the maximum PNS mass is different for each model
and EOS.  Models evolved with the LS180 EOS have PNSs with maximum
baryonic masses ranging from $2.1$ to $2.5\,M_\odot$. Models with low
$\xi_{2.5}$ correspond to the lower end of this mass range. For these
models, $t_\mathrm{BH}$ can be somewhat less than the free fall time
of the 2.5 $M_\odot$ mass element, because less material is needed to
form a BH.  The maximum baryonic PNS mass range for models using the
LS220 EOS is somewhat higher, $2.3 - 2.6\,M_\odot$. BH formation times
for these models are more in line with the Newtonian free fall
prediction. Models evolved with the LS375 and HShen EOS have PNSs that
must accrete upward of $\sim 3\,M_\odot$ of material before becoming
unstable. This significantly increases $t_\mathrm{BH}$ above the free
fall prediction for the $\xi_{2.5}$ characteristic mass element.

\begin{figure}[t]
\begin{center}
\includegraphics[width=\columnwidth]{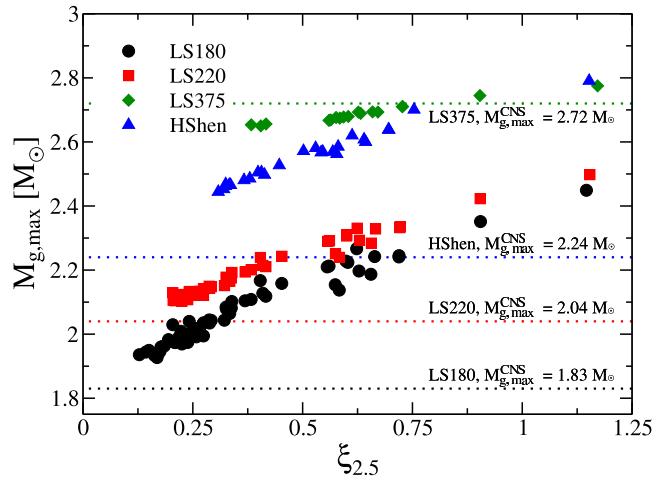} 
\caption{Maximum gravitational PNS masses as a function of the bounce
  compactness ($\xi_{2.5}$) for all nonrotating models presented in
  Table~\ref{tab:results} that form BHs within $3.5\,$s after bounce.
  Simulations performed with the LS180, LS220, LS375, and HShen EOS
  are labeled with circles, squares, diamonds, and triangles,
  respectively.  Also shown (dotted lines, labeled) are the maximum
  gravitational cold-neutron star (CNS) masses,
  $M_\mathrm{g,\,max}^\mathrm{CNS}$, numerical values are 1.83, 2.04,
  2.72, and $2.24\,M_\odot$ for the LS180, LS220, LS375, and HShen EOS,
  respectively.}\label{fig:mgrav_vs_moverr}
\end{center}
\end{figure}

Thermal pressure support can increase the maximum gravitational PNS
mass ($M_{\mathrm{g,\,max}}$) as we have seen in
Section \ref{sec:eosdependence} for the $s$40WH07 model. In
Figure~\ref{fig:mgrav_vs_moverr}, we plot $M_{\mathrm{g,\,max}}$ as a
function of $\xi_{2.5}$ for all nonrotating models listed in
Table~\ref{tab:results}. As obvious from this figure,
$M_{\mathrm{g,\,max}}$ depends in a predictable way not only on the
EOS, but also on the bounce compactness of the presupernova model.
Progenitors with high $\xi_{2.5}$, in addition to forming BHs faster,
create PNSs that are stable to higher masses. This is a simple
consequence of the fact that progenitors with larger $\xi_{2.5}$ have
iron cores with systematically higher entropies and masses
significantly above the cold Chandrasekhar mass (see
Table~\ref{tab:initialmodels} and \citealt{baron:90} and
\citealt{burrows:83}).  Adiabatic collapse leads to higher PNS
temperatures after bounce in progenitors with high $\xi_{2.5}$ and,
hence, more thermal support.  This leads to higher maximum PNS masses.
This effect can be large, up to $25\%$ for models with large
$\xi_{2.5}$ and soft EOS.

\subsection{Preventing BH Formation with Artificial Neutrino-driven
Explosions}
\label{sec:lums}

In a successful CCSN, the shock is reenergized before enough material
can accrete onto the PNS to make it unstable. While fully
self-consistent spherically symmetric simulations generally fail to
explode in all but a few very low mass progenitors
(cf. \citealt{kitaura:06,burrows:07c}), one can explode any star by the 1D
neutrino mechanism by artificially increasing the energy deposition in
the postshock region. Without such an increase, all of our simulations
fail to explode. Our parameterized heating ($f_\mathrm{heat}$ in
Equation~\ref{eq:heating}) allows us to explore ``how much'' neutrino
heating is needed to explode a given model (in 1D).  By comparison
with results from previous self-consistent radiation hydrodynamics
simulations we can then estimate whether a given progenitor and EOS
combination is more likely to lead to an explosion or BH formation.

Our method for driving explosions is similar to \cite{murphy:08}, but
has the advantage of being proportional to the neutrino luminosity
obtained from the neutrino leakage scheme and therefore conserves
energy. We iteratively determine the critical value of
$f_\mathrm{heat}$ to within $1\,$\% to what is needed for a successful
explosion for a large subset of our models and the LS180, LS220, and
HShen EOS. Of particular interest in this analysis is the
time-averaged heating efficiency of the critical model,
$\bar{\eta}^\mathrm{crit}_\mathrm{heat}$.  We define
$\bar{\eta}_\mathrm{heat}$ as
\begin{equation} \bar{\eta}_\mathrm{heat} =
  \overline{\int_{\mathrm{gain}}
    \dot{q}_\nu^+\, dV \bigg / \left(L_{\nu_e} +
      L_{\bar{\nu}_e}\right)_{r_\mathrm{gain}}}\,,\label{eq:heateff}
\end{equation}
where $\dot{q}_\nu^+$ is the net energy deposition rate and the
neutrino luminosities are taken at the gain radius. We perform the
time average between bounce and explosion, the latter time defined as
when the postshock region assumes positive velocities and accretion
onto the PNS ceases. $\bar{\eta}^\mathrm{crit}_\mathrm{heat}$ is a
useful quantity because it characterizes how much of the available
luminosity must be redeposited on average to explode a given
progenitor. This is rather independent of transport scheme and
code. For example, for the 15 $M_\odot$ ZAMS solar-metallicity
progenitor of \cite{ww:95} we find
$\bar{\eta}^\mathrm{crit}_\mathrm{heat} \sim 0.13$. \cite{buras:06a}
who also artificially exploded this progenitor in 1D, though with much
more sophisticated neutrino transport, find\footnote{This we deduce
  from their Figure~28, bottom panel. Note that their $\delta_t
  E_\mathrm{cool}$ includes all neutrinos, not just $\nu_e$ and
  $\bar{\nu}_e$.}  an average heating efficiency of $0.1-0.15$ which
is consistent with our result. Note, however, that \cite{marek:09}
observed in the same progenitor the onset of a self-consistent
neutrino-driven explosion in 2D at an average heating efficiency of
$\sim 0.07$. This indicates a dependence of
$\bar{\eta}^\mathrm{crit}_\mathrm{heat}$ on dimensionality, should be
kept in mind, and is consistent with recent work that suggest that
dimensionality may be the key to successful neutrino-driven explosions
\citep{murphy:08,nordhaus:10}.

\begin{figure}[t]
\begin{center}
\includegraphics[width=\columnwidth]{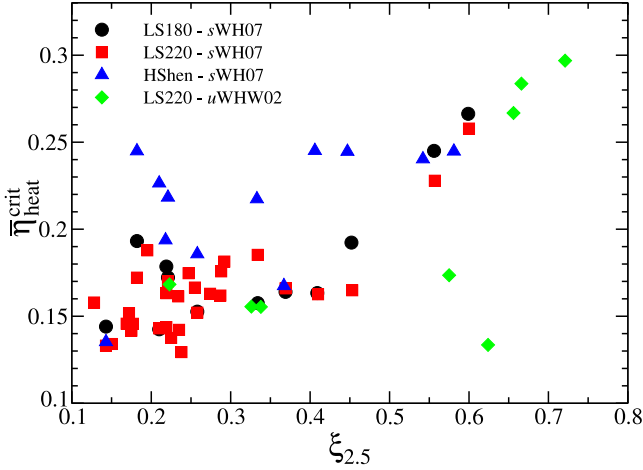} 
\caption{$\bar{\eta}_\mathrm{heat}^\mathrm{crit}$ obtained with
  \code{GR1D} as a function of bounce compactness. Plotted are models
  from the $s$WH07 data set using the LS180, LS220, and HShen EOS; and
  models from the $u$WHW02 data set using the LS220
  EOS. }\label{fig:heating_etah_vs_xi}
\end{center}
\end{figure}

\begin{deluxetable}{cccc|cccc}
\tablecolumns{8}
\tablewidth{0pc}
\tablecaption{Explosion Properties}
\tablehead{
 $M_{\mathrm{ZAMS}}$ & $\xi_{2.5}$ & $f^\mathrm{crit}_\mathrm{heat}$ &
 $\bar{\eta}^\mathrm{crit}_\mathrm{heat}$ & $M_{\mathrm{ZAMS}}$ & $\xi_{2.5}$ & $f^\mathrm{crit}_\mathrm{heat}$ &
 $\bar{\eta}^\mathrm{crit}_\mathrm{heat}$ \\
 $\left(M_\odot\right)$ & & & & $\left(M_\odot\right)$ & & & }
\startdata
\multicolumn{4}{c|}{$s$WH07 LS220} & \multicolumn{4}{|c}{$s$WH07 LS180}\\
\phantom{1}14 & 0.128 & 1.17 & 0.158 & 15 & 0.182 & 1.16 & 0.193\\
\phantom{1}15 & 0.182 & 1.17 & 0.172 & 21 & 0.143 & 1.32 & 0.144\\
\phantom{1}16 & 0.150 & 1.33 & 0.134 & 23 & 0.452 & 1.18 & 0.192\\
\phantom{1}17 & 0.169 & 1.32 & 0.146 & 24 & 0.409 & 1.16 & 0.163\\
\phantom{1}18 & 0.195 & 1.17 & 0.188 & 25 & 0.334 & 1.13 & 0.158\\
\phantom{1}19 & 0.177 & 1.24 & 0.146 & 27 & 0.258 & 1.18 & 0.153\\
\phantom{1}20 & 0.288 & 1.15 & 0.176 & 30 & 0.219 & 1.16 & 0.179\\
\phantom{1}21 & 0.143 & 1.34 & 0.133 & 35 & 0.369 & 1.14 & 0.164\\
\phantom{1}22 & 0.292 & 1.15 & 0.181 & 40 & 0.599 & 1.32 & 0.266\\
\phantom{1}23 & 0.453 & 1.17 & 0.165 & 45 & 0.556 & 1.26 & 0.245\\
\phantom{1}24 & 0.410 & 1.15 & 0.163 & 50 & 0.221 & 1.18 & 0.172\\
\phantom{1}25 & 0.334 & 1.14 & 0.185 & 80 & 0.210 & 1.22 & 0.142\\
\phantom{1}26 & 0.235 & 1.21 & 0.142 & \multicolumn{4}{c}{$s$WH07 HShen}\\
\phantom{1}27 & 0.258 & 1.20 & 0.152 & 15 & 0.182 & 1.30 & 0.245\\
\phantom{1}28 & 0.274 & 1.16 & 0.163 & 21 & 0.143 & 1.50 & 0.135\\
\phantom{1}29 & 0.225 & 1.25 & 0.138 & 23 & 0.447 & 1.27 & 0.245\\
\phantom{1}30 & 0.219 & 1.18 & 0.163 & 24 & 0.406 & 1.31 & 0.245\\
\phantom{1}31 & 0.219 & 1.21 & 0.144 & 25 & 0.333 & 1.49 & 0.217\\
\phantom{1}32 & 0.255 & 1.17 & 0.166 & 27 & 0.258 & 1.52 & 0.186\\
\phantom{1}33 & 0.287 & 1.15 & 0.162 & 30 & 0.218 & 1.32 & 0.194\\
\phantom{1}35 & 0.369 & 1.13 & 0.166 & 35 & 0.367 & 1.37 & 0.167\\
\phantom{1}40 & 0.600 & 1.30 & 0.259 & 40 & 0.581 & 1.22 & 0.245\\
\phantom{1}45 & 0.557 & 1.25 & 0.228 & 45 & 0.542 & 1.24 & 0.240\\
\phantom{1}50 & 0.221 & 1.19 & 0.170 & 50 & 0.221 & 1.41 & 0.218\\
\phantom{1}55 & 0.238 & 1.24 & 0.129 & 80 & 0.210 & 1.50 & 0.226\\
\phantom{1}60 & 0.175 & 1.29 & 0.142 & &&&\\
\phantom{1}70 & 0.234 & 1.21 & 0.161 & \multicolumn{4}{c}{$s$WW95 LS180}\\
\phantom{1}80 & 0.210 & 1.24 & 0.143 & 15 & 0.088 & 1.33 & 0.130\\
 100 & 0.247 & 1.15 & 0.175 & &&&\\
 120 & 0.172 & 1.25 & 0.152 & &&&\\
 \multicolumn{8}{c}{$u$WHW02 LS220}\\
\phantom{1}20 & 0.338 & 1.13 & 0.155 & 40 & 0.721 & 1.44 & 0.297\\
\phantom{1}25 & 0.223 & 1.16 & 0.168 & 45 & 0.656 & 1.22 & 0.267\\
\phantom{1}30 & 0.326 & 1.13 & 0.156 & 50 & 0.575 & 1.09 & 0.174\\
\phantom{1}35 & 0.666 & 1.37 & 0.284 & 60 & 0.624 & 1.12 & 0.133
\enddata
\tablecomments{$f^\mathrm{crit}_\mathrm{heat}$ corresponds to the critical value
 needed to cause a successful explosion in \code{GR1D}.
 $\bar{\eta}^\mathrm{crit}_\mathrm{heat}$ is the associated critical average heating
 efficiency defined in Equation~(\ref{eq:heateff}).}
\label{tab:heatingresults}
\end{deluxetable} 

Since \code{GR1D}'s leakage/heating scheme is only a rough
approximation to true neutrino transport, and because our simulations
assume spherical symmetry, we cannot make very robust quantitative
predictions for any one particular model, but rather study the
collective trends exhibited by the entire set of 62 progenitors that
we consider here. In Figure~\ref{fig:heating_etah_vs_xi}, as a function
of bounce compactness $\xi_{2.5}$, we plot
$\bar{\eta}_\mathrm{heat}^\mathrm{crit}$ for all considered models and
EOS. The data are summarized in Table~\ref{tab:heatingresults}. We can
divide the results into two general regimes: models with $\xi_{2.5}
\lesssim 0.45$ and those with $\xi_{2.5} \gtrsim 0.45$.

For many models with $\xi_{2.5} \lesssim 0.45$, oscillations in the
shock position are ubiquitous near the transition from failing to
exploding supernovae in 1D (cf.  \citealt{murphy:08};
\citealt{buras:06a}; \citealt{fernandez:09b}). For both the LS180
and LS220 EOS, the $\bar{\eta}_\mathrm{heat}^\mathrm{crit}$ required
for an explosion, modulo noise, is roughly constant and $\sim 0.16$ on
average for low $\xi_{2.5}$ models. Hence, explosion is the likely
outcome of core collapse for progenitors with $\xi_{2.5} \lesssim
0.45$ if the nuclear EOS is similar to the LS180 or LS220 case.

The noise in the $\bar{\eta}_\mathrm{heat}^\mathrm{crit}$ distribution
(absolute variations by up to $\sim 10\,$\%) is in part a consequence
of variations in postbounce dynamics, such as the number and duration
of pre-explosion oscillations.  Compositional interfaces in some
progenitor models, where jumps in the density lead to jumps in the
accretion rate, also affect individual models leading to variations in
$\bar{\eta}_\mathrm{heat}^\mathrm{crit}$. For the LS180 and LS220 EOS,
any differences in $\bar{\eta}_\mathrm{heat}^\mathrm{crit}$ with
choice of EOS are indistinguishable given the noise in the data.

For progenitors with $\xi_{2.5} \gtrsim 0.45$, the
$\bar{\eta}_\mathrm{heat}$ required to cause an explosion increases
with $\xi_{2.5}$ when run with the LS180 or LS220 EOS. Progenitors in
this regime have tremendous postbounce accretion rates, accumulating
$\gtrsim 2\,M_\odot$ of baryonic material behind the shock within the
first $\sim 200\,$ms after bounce. Without explosion, they form BHs
within $\lesssim 0.8\,$s (with the LS180 and LS220 EOS).  Hence, a
very high heating efficiency of $\bar{\eta}_\mathrm{heat} \gtrsim
0.23-0.27$ is necessary to drive an explosion at early times against
the huge ram pressure of accretion. It appears unlikely, even when
multi-dimensional dynamics are factored in, that progenitors with $\xi_{2.5}
\gtrsim 0.45$ can be exploded via the neutrino mechanism.  The most
likely outcome of core collapse in such stars is BH formation.

We draw the reader's attention to two outliers in the $u$WHW02 data
set included in Figure~\ref{fig:heating_etah_vs_xi}, the $u$50WHW02 and
$u$60WHW02 progenitors. These models have high $\xi_{2.5}$, but
feature compositional interfaces where the density drops by $\sim
50\,$\%. These are located at a mass coordinate of $1.82\,M_\odot$ and
$2.22\,M_\odot$ in $u$50WHW02 and $u$60WHW02, respectively.  When such
an interface advects through the shock, the accretion rate drops
suddenly, but the core neutrino luminosity remains large and an
explosion is immediately launched. This results in a small value of
$f^\mathrm{crit}_\mathrm{heat}$ and, therefore, in a low required
$\bar{\eta}_\mathrm{heat}$. This demonstrates that the single
parameter $\xi_{2.5}$ is not always sufficient to predict a
progenitor's fate.

In models with $\xi_{2.5} \lesssim 0.45$ and calculated using the
HShen EOS, both $\bar{\eta}_\mathrm{heat}^\mathrm{crit}$ and
$f^\mathrm{crit}_\mathrm{heat}$ are systematically higher than with
the LS180 and LS220 EOS and explosion is less likely. Furthermore, the
qualitative behavior of our simulations is different with the HShen
EOS.  In many models with subcritical $f_\mathrm{heat}$ and
$\bar{\eta}_\mathrm{heat}$, the shock is revived and begins to
propagate to large radii of $\cal{O}$($10^3 - 10^4\,$km), but the
material behind it fails to achieve positive velocities. Hence,
accretion onto the PNS is slowed but does not cease. High values of
$f_\mathrm{heat}$ are needed to avoid this and achieve full
explosions. We caution the reader that this regime may not be well
modeled by our neutrino treatment. Nevertheless, our results suggest
that systematically higher $f_\mathrm{heat}^\mathrm{crit}$ and
$\bar{\eta}_\mathrm{heat}^\mathrm{crit}$ are required to explode
models with the HShen EOS, even at low $\xi_{2.5}$. In contrast to
models using the LS180 or LS220 EOS, models with $\xi_{2.5}
\gtrsim$0.45 with the HShen EOS require roughly constant
$\bar{\eta}_\mathrm{heat}$ to explode.  Since the HShen EOS can
support a high maximum mass, the PNS can withstand BH formation longer
and explosions may set in at later postbounce times when the accretion
rate has dropped sufficiently.

Finally, as an interesting aside, we point out the evolution of the
$u$75WHW02 progenitor evolved with the LS220 EOS. This model has a
bounce compactness of $\sim 1.15$ and, in the absence of an explosion,
forms a BH $\sim 0.285\,$s after bounce (with the LS220 EOS).  This
progenitor has a compositional interface at which the density drops by
$\sim50\,$\% that is located at a baryonic mass coordinate of $\sim
2.5\,M_\odot$.  This is very close to the maximum mass of the
$u$75WHW02 PNS (with the LS220 EOS), and well above the maximum cold
NS (baryonic) mass. The model can be made to explode with
$f_\mathrm{heat}^\mathrm{crit} = 1.35$ with a corresponding
$\bar{\eta}_\mathrm{heat}^\mathrm{crit} = 0.287$. The resulting PNS
has a baryonic (gravitational) mass of $\sim 2.54\,M_\odot$
($2.44\,M_\odot$). Interestingly, within $\sim 100\,$ms after the
launch of the explosion, cooling of the outer PNS layers removes
sufficient thermal pressure, rendering the PNS unstable to collapse
and BH formation. This scenario will necessarily occur within the
cooling phase for any PNS that is initially thermally supported above
the maximum cold NS baryonic mass and is another avenue to BH
formation. In our simulations, this condition is also met only in very
few other models with very high $\xi_{2.5}$ and fairly soft EOS, such
as the 23, 40, and 45 $M_\odot$ progenitors from the $s$WH07 series
using the LS180 EOS. In order to fully investigate this BH formation
channel, a more sophisticated neutrino treatment is required that
allows accurate long-term modeling of PNS cooling \citep{pons:99},
since, in general, the Kelvin-Helmholtz cooling phase of PNS is
$\cal{O}$($10-100\,$s).

\subsection{Connection to Stellar Evolution and ZAMS Conditions}
\label{sec:evolution}

\subsubsection{ZAMS Mass and Metallicity}
\label{sec:ZAMSmass}

Having established the systematic dependence of core collapse and BH
formation on progenitor bounce compactness in Section \ref{sec:prestruct},
we now go further and attempt to connect to the conditions at ZAMS.
Doing this is difficult, and, given the current state and limitations
of stellar evolution theory and modeling, can be done only
approximately.  In general, presupernova stellar structure will depend
not only on initial conditions at ZAMS (mass, metallicity, rotation),
but also on particular evolution history and physics (binary effects,
[rotational] mixing, magnetic fields, nuclear reaction rates, and mass
loss; cf. \citealt{whw:02}).  While keeping this in mind, we limit
ourselves in the following to the exploration of single-star,
nonrotating progenitors without magnetic fields. We
focus on collapse models run with the LS220 EOS, but the general
trends with EOS observed in the previous sections extend to here.

\begin{figure}[t]
\begin{center}
\includegraphics[width=\columnwidth]{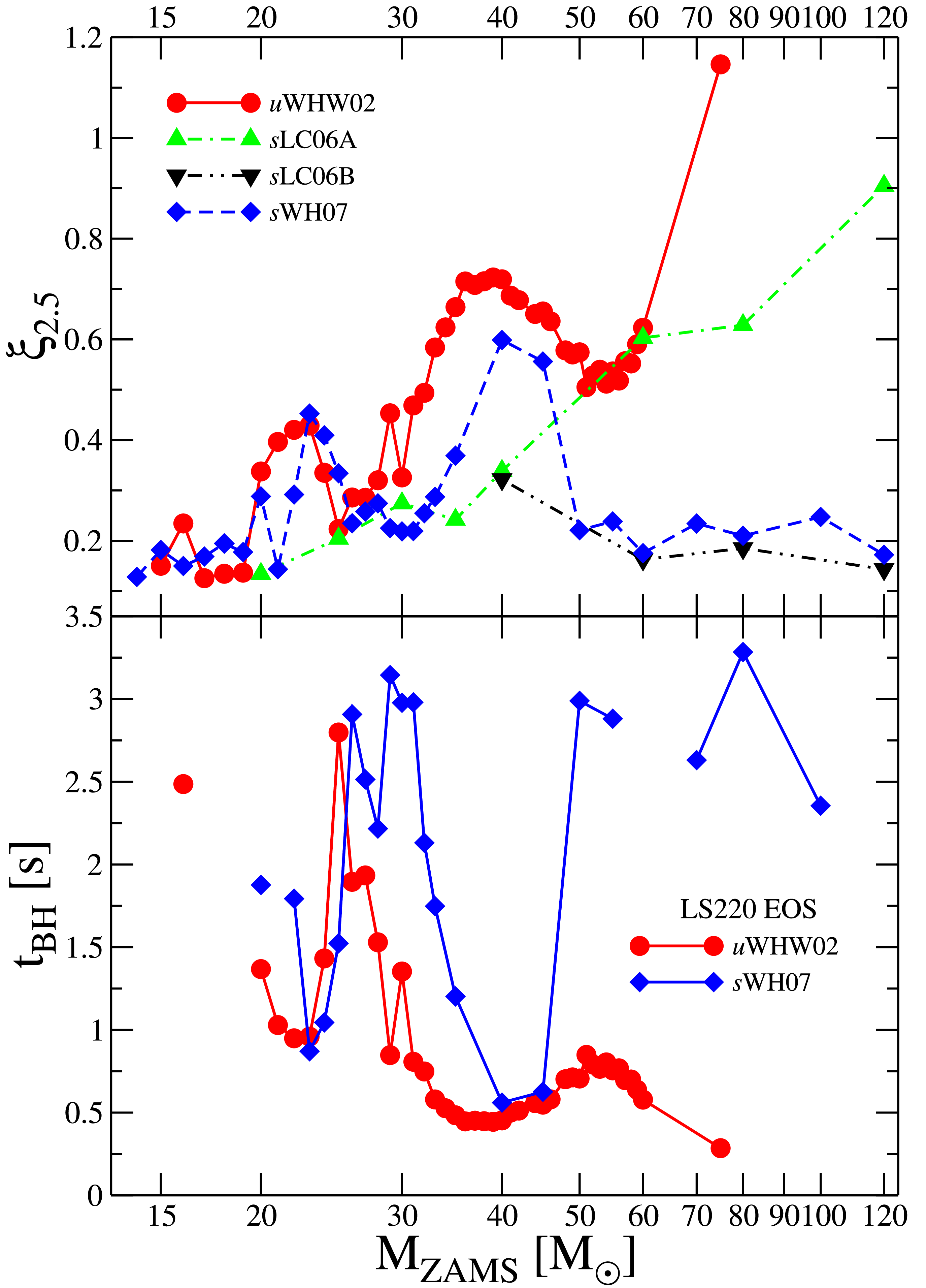} 
\caption{Bounce compactness (top panel) and time to BH
  formation (bottom panel) as functions of ZAMS mass for
  various progenitor sets.  $\xi_{2.5}$ is determined for each model
  at bounce using Equation~(\ref{eq:bouncecompactness}). $t_\mathrm{BH}$ for
  each model is obtained using the LS220 EOS and assuming no
  explosion.  The times to BH formation for progenitor models that
  take longer than $3.5\,$s to form a BH are not shown. Breaks in the
  lines connecting models indicate this. For clarity, the time to BH
  formation is not shown for the $s$LC06A/B series, but is provided in
  Table~\ref{tab:results}.}\label{fig:LS220_tbh}
\end{center}
\end{figure}

In the top panel of Figure~\ref{fig:LS220_tbh}, we plot the bounce
compactness $\xi_{2.5}$ as a function of progenitor ZAMS mass
$M_\mathrm{ZAMS}$ for a range of progenitors from multiple stellar
evolutionary studies.  Even within a given model set, the
$M_\mathrm{ZAMS} - \xi_{2.5}$ mapping is highly non-monotonic.  At the
low end of $M_\mathrm{ZAMS}$ covered by Figure~\ref{fig:LS220_tbh},
where mass loss has little influence even in progenitors of solar
metallicity, variations in $\xi_{2.5}$ are due predominantly to
particularities in late burning stages, caused, e.g., by convective
versus radiative core burning and/or differences in shell burning
episodes \citep{whw:02}. At the high ZAMS-mass end, $\xi_{2.5}$ is
determined by a competition of mass loss and rapidity of
nuclear-burning evolution.

The bottom panel of Figure~\ref{fig:LS220_tbh} depicts the time to BH
formation $t_\mathrm{BH}$ in a failing CCSN as a function of
$M_\mathrm{ZAMS}$ for the $s$WH07 solar-metallicity progenitors of
\cite{woosley:07} and the $u$WHW02 $10^{-4}$ solar-metallicity models
of \cite{whw:02}.  Models of very low $\xi_{2.5}$ that require more
than $3.5\,\mathrm{s}$ to make a BH are omitted. As demonstrated in
Section \ref{sec:prestruct}, $t_\mathrm{BH}$ scales $\propto
(\xi_{2.5})^{-3/2}$ and, hence, progenitors that form BH the fastest
and are (generally, cf. Section \ref{sec:lums}) the hardest to explode are
those with high values of $\xi_{2.5}$.  In the low-metallicity
$u$WHW02 series whose progenitors experience only minuscule mass loss,
BHs form within $\lesssim 1\,\mathrm{s}$ of bounce for
$M_\mathrm{ZAMS} \gtrsim 30\,M_\odot$ and the high bounce compactness
$\xi_{2.5} \gtrsim 0.45$ makes a successful shock revival rather
unlikely (Section \ref{sec:lums}). Hence, the most likely outcome of core
collapse is BH formation in these progenitors. This may also be the
case for $u$WHW02 progenitors in the ZAMS mass range from $\sim
20$ to $25\,M_\odot$.  The $s$WH07 progenitors have high $\xi_{2.5}$ and
form BHs rapidly only in the $M_\mathrm{ZAMS}$ ranges $\sim
23-25\,M_\odot$ and $\sim 35-45\,M_\odot$. At higher ZAMS masses,
strong $O$-star mass loss leads to an early removal of the hydrogen
envelope. Subsequent mass loss in the W-R phase leads to
bare, low-mass, low-compactness carbon oxygen cores in the most
massive progenitors that are unlikely to make BHs.

\subsubsection{Variations with Mass-loss Prescriptions}
\label{sec:masslossprescriptions}

Mass loss is key in determining the observational appearance of a
successful CCSN (e.g., \citealt{filippenko:97, smith:10}), but, as we
have seen in Section \S\ref{sec:ZAMSmass}, also has a
strong effect on presupernova core structure and, thus, on the outcome
of core collapse. The details of mass loss in massive stars are still
rather uncertain (e.g., \citealt{smith:10}), and, unfortunately, there
are few stellar evolution studies that have studied the effects of
variations in mass-loss prescriptions. \cite{limongi:06}\footnote{See
  also \cite{limongi:09} and \cite{meynet:03}} performed such a study,
adopting two different mass-loss rates for the W-R stage of
solar-metallicity stars with $M \gtrsim 40\,M_\odot$.  The $s$LC06B
models are evolved with the W-R mass-loss rates of \cite{langer:89}
that are similar to those used in the $s$WH07 set of
\cite{woosley:07}.  As depicted in the top panel of
Figure~\ref{fig:LS220_tbh}, high-mass $s$LC06B and $s$WH07 models have
similar low $\xi_{2.5}$ and most likely do not form BHs but rather
explode as type-Ibc CCSNe.  The models of the $s$LC06A set were
evolved with the lower (factor of $\sim 2$) W-R mass-loss rates of
\cite{nugis:00}.  The $s$LC06A 60, 80, and 120$M_\odot$ progenitors
have much more mass left at the presupernova stage ($M_\mathrm{pre-SN}
\sim 17-30\,M_\odot$, Figure~\ref{fig:presupernovamass}) and very high
bounce compactness of $\xi_{2.5} \sim 0.6 - 0.9$.  In the likely case
of CCSN failure, a BH forms within $\sim 0.5\,$s with the LS180 EOS
and within $\sim 1.5\,$s for all other EOS.

The above results highlight the sensitivity of outcome predictions on
mass-loss physics and a more solid understanding of this key
ingredient will be necessary to robustly connect ZAMS masses to the
outcome of core collapse for massive stars around and above solar
metallicity.

\subsection{The Formation of Rotating Black Holes}
\label{sec:rotatingBH}

Rotation, if sufficiently rapid, alters the CCSN dynamics via
centrifugal support. This important effect is captured by
\code{GR1D}'s 1.5D rotation treatment, albeit approximately.
Initially, centrifugal support acts to slow the collapse of the inner
core, delaying core bounce. At bounce, lower peak densities are
reached, the hydrodynamic shock forms at a larger radius, and its
enclosed mass is larger. Conservation of angular momentum spins up the
core from precollapse angular velocities that may be of order
rad$\,$s$^{-1}$ to ${\cal{O}}$ ($1000\,\mathrm{rad\,s}^{-1}$) as the
core, initially with $r \sim {\cal{O}}$($1000\,$km), collapses to a
PNS with $r \sim {\cal{O}}$($30\,$km). During the postbounce
evolution, the spinning PNS is stabilized at lower densities, is less
compact, generally colder, and has a softer neutrino spectrum than a
non-spinning counterpart (\citealt{ott:08} and references therein).

\subsubsection{Models with Parameterized Rotation}
\label{sec:parameterizedrotation}

We investigate the effect of rotation in failing CCSNe by assigning
specific angular momentum profiles to the $u$WHW02 model set (see
Section \ref{sec:initialmodels}) via Equation~(\ref{eq:specificj}). This rotation
law approximates what is generally found in stellar evolution
calculations that account for rotation (\citealt{heger:00}; see
\citealt{ott:06spin} for comparison plots). The inner iron core ($\sim
1\,M_\odot$) is rotating almost uniformly. Outside of this core, the
angular velocity drops roughly $ \propto r^{-2}$. In
Table~\ref{tab:rotatingresults}, we summarize key parameters of our
rotating model set. Among them is $T/|W|$, the ratio of rotational
kinetic energy to gravitational binding energy. It is particularly
indicative of the dynamical relevance of rotation.

\begin{figure}[t]
\begin{center}
\includegraphics[width=\columnwidth]{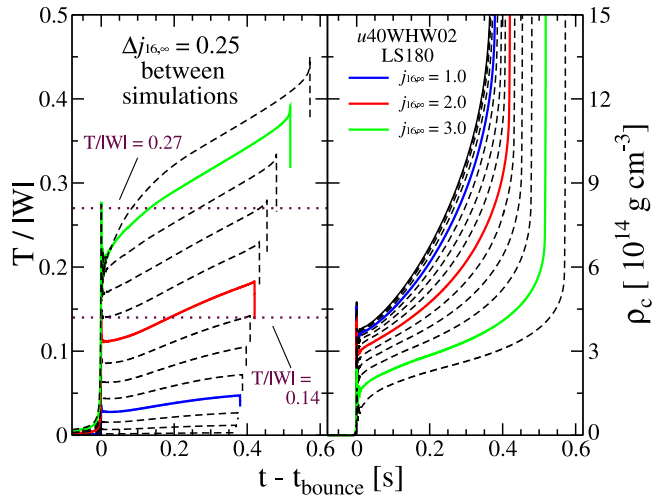} 
\caption{$T/|W|$ (left) and central density ($\rho_c$) (right) during
  the postbounce evolution of the $u$40WHW02 model using the LS180 EOS
  and 14 different initial specific angular momentum profiles.  We
  vary $j_{16,\infty}$ from 0 to 3.25 in increments of 0.25.  For
  clarity we highlight with solid lines the simulations with integer
  values of $j_{16,\infty}$. Lines at $T/|W| = 0.27$ and 0.14 are
  added to denote the dynamical and secular rotational instability
  thresholds.}\label{fig:u40WHW02_rot_rhoc_ToverW}
\end{center}
\end{figure}

In the right panel of Figure~\ref{fig:u40WHW02_rot_rhoc_ToverW}, we
plot the central density evolution of model $u$40WHW02 run with the
LS180 EOS for $j_{16,\infty}$ ranging from 0 to 3.25 in increments of
0.25. While we choose the $u$40WHW02 model here, the results are
generic and apply to all progenitors. $A_{M_\odot}$, of
Equation~(\ref{eq:specificj}), is $936\,$km for this model and the
initial central rotation rate is $1.14\times
j_{16,\infty}\,$rad$\,$s$^{-1}$. The nonrotating model takes $\sim
433\,$ms to reach bounce and a further $\sim 369\,$ms before the PNS
becomes unstable to collapse to a BH with a gravitational mass of
$2.24\,M_\odot$. For the $j_{16,\infty}=1$, 2, and 3 models,
respectively, the times to bounce are 11$\,$ms, 47$\,$ms, and
125$\,$ms greater than in the nonrotating case. Their times to BH
formation $t_\mathrm{BH}$ are 12$\,$ms, 52$\,$ms, and 150$\,$ms longer
than in the nonrotating case.  The maximum gravitational PNS masses
$M_\mathrm{g,\,\mathrm{max}}$ are $0.03\,M_\odot$, $0.09\,M_\odot$,
and $0.28\,M_\odot$ greater. We find that the time to bounce, time to
BH formation, and maximum gravitational PNS mass increase above the
nonrotating values proportional to $\sim (j_{16,\infty})^2$. The
increase in $t_\mathrm{BH}$ is due almost entirely to the increase in
$M_{g,\,\mathrm{max}}$, since the accretion rate is not significantly
affected by rotation.

The lower temperatures and densities of rotating PNSs lead to
systematically lower mean neutrino energies and total radiated energy
from the PNS core (time-averaged total luminosities are summarized in
Table~\ref{tab:rotatingresults}).  \cite{fh:00} and \cite{ott:08}, who
considered similarly rapidly rotating models, also see this effect.
There is a clear trend toward lower $L_\nu$ with increasing
$j_{16,\infty}$ and for a given model and at a given time, with
increasing $j_{16,\infty}$, less gravitational binding energy has been
carried away by neutrinos and $M_{g}$ is larger. Given essentially
unaltered accretion rates, one may expect earlier PNS collapse and BH
formation. This, however, is not the
  case in models run with the LS180, LS220, and HShen EOS, since the
  centrifugally increased $M_{g,\mathrm{max}}$ systematically
  outweighs the increased gravitational mass due to the lowered
  neutrino emission. For these EOS, the time to BH formation is
  delayed by rotation.  For models run with the
extremely stiff LS357 EOS the situation is different. For them, the
centrifugal support provided by rotation is too weak to significantly
enhance $M_{g,\mathrm{max}}$ and, hence, BHs form faster with
increasing $j_{16,\infty}$.

In the left panel of Figure~\ref{fig:u40WHW02_rot_rhoc_ToverW}, we
plot the $T/|W|$ evolution for the rotating $u$40WHW02 model series
run with the LS180 EOS. During collapse, gravitational binding energy
is transferred to rotational energy, increasing the value of
$T/|W|$. Similar to how the central density overshoots its new
equilibrium, $T/|W|$ also exhibits a local maximum at
bounce. Continued accretion and contraction of the PNS increases
$T/|W|$ throughout the postbounce evolution for all models.  Initially
very rapidly spinning models experience core bounce under the strong
influence of centrifugal effects, leading to reduced compactness and
$T/|W|$ at bounce.  These qualitative features are in good agreement
with what was found by previous extensive parameter studies of
rotating core collapse
\citep{ott:06spin,dimmelmeier:08}. Quantitatively, we find and
summarize in Table~\ref{tab:rotatingresults} that models with
$j_{16,\infty} \lesssim 1.5$ yield $T/|W| \lesssim 0.14$ throughout
their entire evolution. Models with $j_{16,\infty} \gtrsim 2.25$ have
$T/|W| \gtrsim 0.14$ during their entire postbounce evolution.  Models
that have $j_{16,\infty} \lesssim 2.25$ have $T/|W| \lesssim 0.27$ at
all times. Models with $j_{16,\infty} \gtrsim 2.5$ reach $T/|W|
\gtrsim 0.27$ before BH formation. When considering these numbers, it
is important to keep in mind that \code{GR1D}'s 1.5D approach to
rotation has the tendency to overestimate $T/|W|$ in rapidly spinning
models. \cite{ott:06spin} found model-dependent differences in $T/|W|$
of $\mathcal{O}(10\,\%)$ between 1.5D and 2D.  In addition, GR1D's
neutrino leakage scheme also tends to lead to somewhat more compact
PNS cores and consequently higher $T/|W|$ than would be expected from
full neutrino transport calculations

The systematics of $T/|W|$ depicted by
Figure~\ref{fig:u40WHW02_rot_rhoc_ToverW} (left) and listed in
Table~\ref{tab:rotatingresults}, albeit only approximate due to
\code{GR1D}'s 1.5D treatment of rotation, shed interesting light on
the potential role of nonaxisymmetric rotational instabilities during
the evolution of failing CCSNe. Of course, due to its 1.5D nature,
\code{GR1D} cannot track the development of such multi-dimensional
dynamics. Analytic theory and to some extent 3D computational modeling
have identified multiple instabilities that may lead to triaxial
deformation of PNSs, redistribution of angular momentum, and to the
radiation of rotational energy and angular momentum via gravitational
waves (see \citealt{stergioulas:03} and \citealt{ott:09} for
reviews). A global dynamical instability sets in for $T/|W| \gtrsim
0.27$ \citep{chandrasekhar69c}, leading to a lowest-order $m$ = 2
``bar'' deformation. Global secular instability, driven by viscosity or
GW back-reaction sets in at $T/|W| \gtrsim 0.14$
(\citealt{chandrasekhar:70, friedman:78}). Finally, dynamical shear
instabilities, arising as a result of differential rotation, may lead
to partial or global nonaxisymmetric deformation at even lower values
of $T/|W|$ ($\gtrsim 0.05$; e.g.,~\citealt{saijo:03, ott:07prl,
  scheidegger:08, corvino:10}, and references therein). In nature, and
in full 3D simulations, these instabilities, through gravitational
radiation or redistribution of angular momentum, will effectively and
robustly prevent $T/|W|$ from surpassing the corresponding $T/|W|$
threshold.  The growth times of dynamical instabilities are short,
$\cal{O}$(ms).  Secular instabilities grow on timescales set by the
driving process and are typically $\cal{O}$(s) \citep{lai:95}. The
low-$T/|W|$ shear instabilities in PNSs appear to grow on intermediate
timescales of $\cal{O}$(10-100$\,$ms)
(e.g., \citealt{ott:07prl,scheidegger:08}).

\begin{figure}[t]
\begin{center}
\includegraphics[width=\columnwidth]{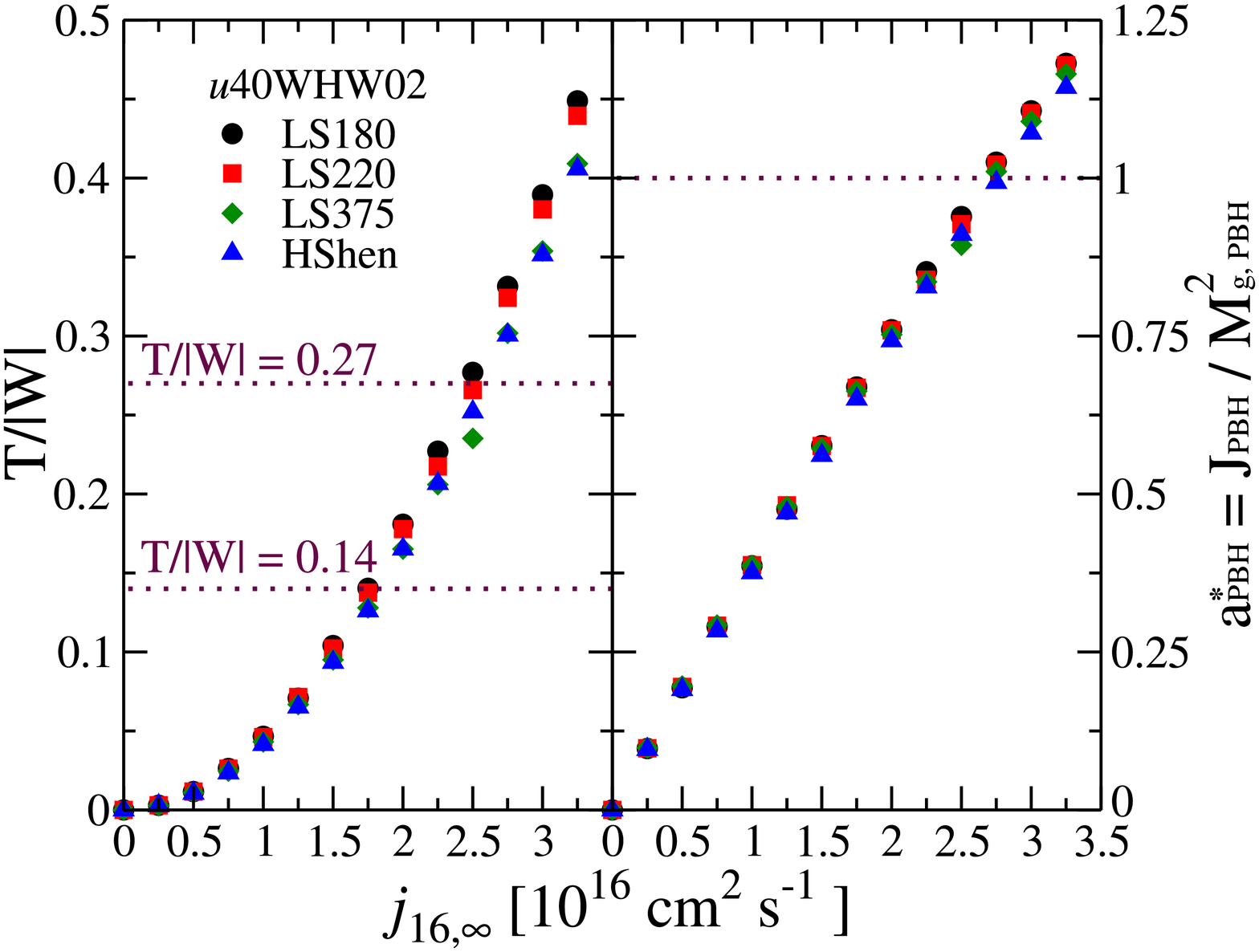} 
\caption{Left: $T/|W|$ for a range of initial $j_{16,\ \infty}$
  and EOS for the $u$40WHW02 progenitor.  We denote the value of
  $T/|W|$ thresholds for the dynamical rotational instability,
  $T/|W|_\mathrm{dyn} =0.27$, and the secular instability, $T/|W|_\mathrm{sec} =
  0.14$.  Right: dimensionless spin parameter
  $a^*_\mathrm{PBH}$ for the PNS at the last stable configuration
  prior to collapse to a BH.  Note that $a^*_\mathrm{PBH} > 1$ is
  generally allowed by GR but a BH must have $a^* < 1$.  PNSs that
  could reach $a^*_\mathrm{PBH} > 1$ are nonaxisymmetrically unstable and
  will be limited to $a^*_\mathrm{PBH}$ below 1.  For the $u$WHW02
  model, the initial central rotation rate is given as $\Omega_c = 1.141 \times
  j_{16,\ \infty}\,$rad$\,$s$^{-1}$.}\label{fig:a_and_ToverW}
\end{center}
\end{figure}

In Figure~\ref{fig:a_and_ToverW}, we plot the value of $T/|W|$ (left
panel) and the dimensionless spin of the protoblack hole (PBH),
$a^*_\mathrm{PBH} = J_\mathrm{PBH} / (M_\mathrm{g,\,PBH}^2)$ (right
panel) at the onset of BH formation (when $\alpha_c = 0.3$) for the
same values of $j_{16,\infty}$ used in
Figure~\ref{fig:u40WHW02_rot_rhoc_ToverW}. Assuming that the entire PNS
is promptly swallowed once the horizon appears, $a^*_\mathrm{PBH}$
corresponds to the BH birth spin\footnote{Note that this may not
  necessarily be what happens. Outer PNS material may become
  centrifugally supported, falling into the nascent BH only on an
  accretion timescale \citep{duez:06b}.}.  We again show results for
model $u$40WHW02, but for all four EOS. The data are also presented in
Table~\ref{tab:rotatingresults} for these and other models.  $T/|W|$
at BH formation scales $\propto (j_{16,\infty})^2$:
$T/|W|_\mathrm{PBH}$ is $\sim 0.05$, $\sim 0.1$, $\sim 0.2$, and
$\sim 0.3$ at $j_{16,\infty}$ of $\sim 1$, $\sim 1.5$, $\sim 2.2$, and
$\sim 2.75$, respectively. $a^*_\mathrm{PBH}$ scales linearly with
$j_{16,\infty}$, reaching a maximally Kerr value of $a^*_\mathrm{PBH}
\sim 1$ at $j_{16,\infty} \sim 2.75$. $T/|W|_\mathrm{PBH}$ and
$a^*_\mathrm{PBH}$ vary little with EOS.

A disturbing fact depicted by Figure~\ref{fig:a_and_ToverW} is that our
1.5D simulations predict BH birth spins $a^* \gtrsim 1$ for
$j_{16,\infty} \gtrsim 2.75$. In Kerr theory, such BHs cannot exist
with a horizon. They would instead be naked singularities, violating
the cosmic censorship conjecture \citep{penrose:69}. However, when
comparing right and left panels of Figure~\ref{fig:a_and_ToverW}, one
notes that all models achieving $a^* \gtrsim 1$ are predicted to reach
$T/|W|$ in excess of $0.27$. Hence, in nature and in a 3D simulation,
these PNS will be dynamically nonaxisymmetrically unstable and angular
momentum redistribution and gravitational radiation will limit their
$T/|W|$ robustly below $\sim 0.27$, corresponding to $a^* \lesssim
0.9$. Rotational instabilities at lower values of $T/|W|$ may also be
relevant.  Dynamical shear instabilities have timescales significantly
less than the time to BH formation. Secular rotational instabilities
may be relevant if the true nuclear EOS allows for a large maximum PNS
mass as more time is needed to accrete the necessary material to form
a BH (see Section \ref{sec:eosdependence}). Large $t_\mathrm{BH}$ is also
possible if $\xi_{2.5}$ is small (see Section \ref{sec:prestruct}) therefore
allowing secular instabilities to grow. In all rotating models
considered here (see Table~\ref{tab:rotatingresults}), PNSs stable
against the dynamical rotational instability with $T/|W| \lesssim
0.25-0.27$ throughout their postbounce evolution have
$a^*_\mathrm{PBH} \lesssim 0.9$. Similarly, PNSs with $T/|W| \lesssim
0.14-0.16$, the threshold for secular instability, have
$a^*_\mathrm{PBH} \lesssim 0.6-0.7$. If low-$T/|W|$ instabilities are
effective at limiting $T/|W|$ in PNSs on short timescales, nascent BH
spins may be limited to low values ($a^* \lesssim 0.4$ for a $T/|W|$
instability threshold of $\sim 0.05$).

\begin{deluxetable*}{rccccccccccc}
\tablecolumns{12}
\tablewidth{0pc} 
\tablecaption{Properties of Rotating Models}
\tablehead{ Model & $\Omega_\mathrm{c,\,init}$\tablenotemark{a} &
  $J$\tablenotemark{b} & $T/|W|_\mathrm{init}$\tablenotemark{c} &
  $T/|W|_\mathrm{bounce}$\tablenotemark{d} & $t_\mathrm{BH} $ &
  $M_\mathrm{b,max} $& $M_\mathrm{g,max} $& $J_\mathrm{PBH}$\tablenotemark{e} &
  $T/|W|_\mathrm{PBH}$\tablenotemark{f} &
  $a^{*}_\mathrm{PBH}$\tablenotemark{g} & $\overline{L_\nu}$\tablenotemark{h}\\
  & (rad$\,$s$^{-1}$) & (10$^{49}\,$erg$\,$s) & (\%) & (\%) & (s) &
  ($M_\odot$) & ($M_\odot$) & (10$^{49}\,$erg$\,$s) & (\%) & & (100
  B$\,$s$^{-1}$) } 
\startdata
$u$30WHW02LS180J1.0 & 1.34 & \phantom{00}9.97 &  0.048 & \phantom{0}3.32 & 0.990 & 2.24 & 2.11 & 1.61 & \phantom{0}5.06 & 0.41 & 2.07 \\
$u$30WHW02LS180J2.0 & 2.69 & \phantom{0}19.94 &  0.193 & \phantom{}13.44 & 1.223 & 2.33 & 2.20 & 3.48 & \phantom{}20.08 & 0.81 & 1.58 \\
$u$40WHW02LS180J0.5 & 0.57 & \phantom{00}8.07 &  0.006 & \phantom{0}0.87 & 0.371 & 2.35 & 2.25 & 0.86 & \phantom{0}1.16 & 0.19 & 4.99 \\
$u$40WHW02LS180J1.0 & 1.14 & \phantom{0}16.14 &  0.025 & \phantom{0}3.50 & 0.381 & 2.37 & 2.27 & 1.75 & \phantom{0}4.66 & 0.39 & 4.77 \\
$u$40WHW02LS180J1.5 & 1.71 & \phantom{0}24.21 &  0.057 & \phantom{0}7.96 & 0.398 & 2.41 & 2.30 & 2.69 & \phantom{}10.41 & 0.58 & 4.41 \\
$u$40WHW02LS180J2.0 & 2.28 & \phantom{0}32.29 &  0.101 & \phantom{}14.26 & 0.421 & 2.45 & 2.35 & 3.69 & \phantom{}18.09 & 0.76 & 3.95 \\
$u$40WHW02LS180J2.5 & 2.85 & \phantom{0}40.36 &  0.157 & \phantom{}21.91 & 0.455 & 2.52 & 2.41 & 4.81 & \phantom{}27.70 & 0.94 & 3.46 \\
$u$40WHW02LS180J3.0 & 3.42 & \phantom{0}48.43 &  0.226 & \phantom{}24.92 & 0.519 & 2.63 & 2.52 & 6.19 & \phantom{}38.94 & [1.11] & 2.93 \\
$u$50WHW02LS180J1.0 & 1.31 & \phantom{0}22.51 &  0.020 & \phantom{0}3.05 & 0.588 & 2.30 & 2.18 & 1.66 & \phantom{0}4.79 & 0.40 & 3.30 \\
$u$50WHW02LS180J2.0 & 2.62 & \phantom{0}45.02 &  0.079 & \phantom{}12.58 & 0.662 & 2.38 & 2.26 & 3.54 & \phantom{}18.80 & 0.78 & 2.73 \\
$u$60WHW02LS180J1.0 & 1.05 & \phantom{0}30.03 &  0.013 & \phantom{0}3.00 & 0.453 & 2.38 & 2.28 & 1.72 & \phantom{0}4.36 & 0.38 & 3.64 \\
$u$60WHW02LS180J2.0 & 2.10 & \phantom{0}60.07 &  0.052 & \phantom{}12.46 & 0.540 & 2.44 & 2.34 & 3.60 & \phantom{}17.17 & 0.75 & 2.92 \\
\hline
$u$30WHW02LS220J1.0 & 1.34 & \phantom{00}9.97 &  0.048 & \phantom{0}3.30 & 1.419 & 2.38 & 2.20 & 1.75 & \phantom{0}4.97 & 0.41 & 1.89 \\
$u$30WHW02LS220J2.0 & 2.69 & \phantom{0}19.94 &  0.193 & \phantom{}13.43 & 1.697 & 2.45 & 2.29 & 3.77 & \phantom{}19.89 & 0.82 & 1.41 \\
$u$40WHW02LS220J0.5 & 0.57 & \phantom{00}8.07 &  0.006 & \phantom{0}0.86 & 0.455 & 2.47 & 2.34 & 0.94 & \phantom{0}1.16 & 0.19 & 5.19 \\
$u$40WHW02LS220J1.0 & 1.14 & \phantom{0}16.14 &  0.025 & \phantom{0}3.48 & 0.462 & 2.49 & 2.36 & 1.89 & \phantom{0} 4.61 & 0.39 & 4.95 \\
$u$40WHW02LS220J1.5 & 1.71 & \phantom{0}24.22 &  0.057 & \phantom{0}7.94 & 0.474 & 2.51 & 2.39 & 2.89 & \phantom{}10.22 & 0.58 & 4.58 \\
$u$40WHW02LS220J2.0 & 2.28 & \phantom{0}32.29 &  0.100 & \phantom{}14.29 & 0.488 & 2.55 & 2.42 & 3.93 & \phantom{}17.78 & 0.76 & 4.09 \\
$u$40WHW02LS220J2.5 & 2.85 & \phantom{0}40.36 &  0.157 & \phantom{}22.13 & 0.508 & 2.59 & 2.48 & 5.00 & \phantom{}26.58 & 0.93 & 3.57 \\
$u$40WHW02LS220J3.0 & 3.42 & \phantom{0}48.43 &  0.226 & \phantom{}24.57 & 0.549 & 2.67 & 2.56 & 6.36 & \phantom{}38.01 & [1.10] & 2.99 \\
$u$50WHW02LS220J1.0 & 1.31 & \phantom{0}22.51 &  0.020 & \phantom{0}3.04 & 0.725 & 2.43 & 2.28 & 1.82 & \phantom{0}4.73 & 0.40 & 3.45 \\
$u$50WHW02LS220J2.0 & 2.62 & \phantom{0}45.03 &  0.079 & \phantom{}12.60 & 0.777 & 2.49 & 2.35 & 3.81 & \phantom{}18.51 & 0.78 & 2.86 \\
$u$60WHW02LS220J1.0 & 1.05 & \phantom{0}30.03 &  0.013 & \phantom{0}2.99 & 0.602 & 2.48 & 2.35 & 1.84 & \phantom{0}4.31 & 0.38 & 3.48 \\
$u$60WHW02LS220J2.0 & 2.10 & \phantom{0}60.07 &  0.052 & \phantom{}12.49 & 0.664 & 2.53 & 2.41 & 3.80 & \phantom{}16.83 & 0.74 & 2.87 \\
\hline
$u$30WHW02LS375J1.0 & 1.34 & \phantom{00}9.97 &  0.048 & \phantom{0}3.24 & \nodata & (2.81) & (2.50) &  (2.30) & \phantom{0}(5.14) & (0.42) & (1.35) \\
$u$30WHW02LS375J2.0 & 2.69 & \phantom{0}19.94 &  0.192 & \phantom{}13.32 & \nodata & (2.82) & (2.55) &  (4.55) & (18.73) & (0.79) & (1.15) \\
$u$40WHW02LS375J0.5 & 0.57 & \phantom{00}8.07 &  0.006 & \phantom{0}0.85 & 0.941 & 3.02 & 2.71 & 1.27 & \phantom{0}1.10 & 0.20 & 5.53 \\
$u$40WHW02LS375J1.0 & 1.14 & \phantom{0}16.14 &  0.025 & \phantom{0}3.44 & 0.926 & 3.02 & 2.72 & 2.52 & \phantom{0}4.31 & 0.39 & 5.34 \\
$u$40WHW02LS375J1.5 & 1.71 & \phantom{0}24.22 &  0.057 & \phantom{0}7.88 & 0.904 & 3.01 & 2.74 & 3.77 & \phantom{0}9.50 & 0.57 & 5.02 \\
$u$40WHW02LS375J2.0 & 2.28 & \phantom{0}32.29 &  0.100 & \phantom{}14.26 & 0.873 & 2.99 & 2.75 & 5.02 & \phantom{}16.52 & 0.75 & 4.59 \\
$u$40WHW02LS375J2.5 & 2.85 & \phantom{0}40.36 &  0.157 & \phantom{}22.36 & 0.845 & 2.99 & 2.78 & 6.07 & \phantom{}23.52 & 0.89 & 4.11 \\
$u$40WHW02LS375J3.0 & 3.42 & \phantom{0}48.43 &  0.226 & \phantom{}23.43 & 0.788 & 2.96 & 2.79 & 7.46 & \phantom{}35.38 & [1.09] & 3.45 \\
$u$50WHW02LS375J1.0 & 1.31 & \phantom{0}22.51 &  0.020 & \phantom{0}2.99 & 1.346 & 3.02 & 2.69 & 2.56 & \phantom{0}4.53 & 0.40 & 4.07 \\
$u$50WHW02LS375J2.0 & 2.62 & \phantom{0}45.03 &  0.079 & \phantom{}12.54 & 1.296 & 2.99 & 2.72 & 5.09 & \phantom{}17.49 & 0.78 & 3.44 \\
$u$60WHW02LS375J1.0 & 1.05 & \phantom{0}30.04 &  0.013 & \phantom{0}2.95 & 1.330 & 3.00 & 2.71 & 2.44 & \phantom{0}4.08 & 0.38 & 3.62 \\
$u$60WHW02LS375J2.0 & 2.10 & \phantom{0}60.07 &  0.052 & \phantom{}12.47 & 1.284 & 2.98 & 2.74 & 4.93 & \phantom{}16.01 & 0.75 & 3.09 \\
\hline
$u$30WHW02HShenJ1.0 & 1.34 & \phantom{00}9.96 &  0.049 & \phantom{0}3.03 & 3.335 & 2.75 & 2.49 & 2.00 & \phantom{0}3.81 & 0.37 & 1.25 \\
$u$30WHW02HShenJ2.0 & 2.69 & \phantom{0}19.92 &  0.195 & \phantom{}12.45 & \nodata & (2.79) & (2.56) &  (4.48) & (\phantom{}18.10) & (0.78) & (1.03) \\
$u$40WHW02HShenJ0.5 & 0.57 & \phantom{00}8.07 &  0.006 & \phantom{0}0.79 & 0.854 & 2.88 & 2.64 & 1.17 & \phantom{0}1.06 & 0.19 & 4.44 \\
$u$40WHW02HShenJ1.0 & 1.14 & \phantom{0}16.13 &  0.025 & \phantom{0}3.21 & 0.873 & 2.90 & 2.67 & 2.36 & \phantom{0}4.13 & 0.38 & 4.28 \\
$u$40WHW02HShenJ1.5 & 1.71 & \phantom{0}24.20 &  0.057 & \phantom{0}7.33 & 0.901 & 2.93 & 2.71 & 3.64 & \phantom{0}9.35 & 0.56 & 4.03 \\
$u$40WHW02HShenJ2.0 & 2.28 & \phantom{0}32.26 &  0.101 & \phantom{}13.27 & 0.932 & 2.97 & 2.76 & 5.00 & \phantom{}16.51 & 0.74 & 3.69 \\
$u$40WHW02HShenJ2.5 & 2.85 & \phantom{0}40.33 &  0.157 & \phantom{}20.79 & 0.959 & 3.01 & 2.82 & 6.36 & \phantom{}25.20 & 0.91 & 3.28 \\
$u$40WHW02HShenJ3.0 & 3.42 & \phantom{0}48.39 &  0.226 & \phantom{}25.09 & 0.999 & 3.06 & 2.88 & 7.83 & \phantom{}35.14 & [1.07] & 2.80 \\
$u$50WHW02HShenJ1.0 & 1.31 & \phantom{0}22.49 &  0.020 & \phantom{0}2.78 & 1.195 & 2.85 & 2.61 & 2.35 & \phantom{0} 4.45 & 0.39 & 3.29 \\
$u$50WHW02HShenJ2.0 & 2.62 & \phantom{0}44.99 &  0.080 & \phantom{}11.62 & 1.266 & 2.93 & 2.71 & 4.93 & \phantom{}17.23 & 0.76 & 2.86 \\
$u$60WHW02HShenJ1.0 & 1.05 & \phantom{0}30.01 &  0.013 & \phantom{0}2.72 & 1.197 & 2.88 & 2.65 & 2.31 & \phantom{0}4.05 & 0.37 & 3.02 \\
$u$60WHW02HShenJ2.0 & 2.10 & \phantom{0}60.02 &  0.052 & \phantom{}11.48 & 1.273 & 2.94 & 2.73 & 4.75 & \phantom{}15.38 & 0.72 & 2.65 \\
\hline
m35OCWH06LS180 & 1.98 & 780.36 &  0.281 & \phantom{0}8.73 & 0.749 & 2.33 & 2.22 & 2.64 & \phantom{}11.68 & 0.61 & 2.37 \\
m35OCWH06LS220 & 1.98 & 780.38 &  0.281 & \phantom{0}8.71 & 0.972 & 2.43 & 2.29 & 2.69 & \phantom{}10.18 & 0.58 & 2.33 \\
m35OCWH06LS375 & 1.98 & 780.42 &  0.281 & \phantom{0}8.66 & 2.194 & 3.01 & 2.69 & 3.47 & \phantom{0}8.34 & 0.54 & 2.36 \\
m35OCWH06HShen & 1.98 & 779.77 &  0.281 & \phantom{0}8.00 & 1.907 & 2.84 & 2.60 & 3.30 & \phantom{0}8.85 & 0.55 & 2.01 \\
E20HLW00LS180 & 3.13 & \phantom{0}18.15 &  0.242 & \phantom{}17.59 & 1.126 & 2.25 & 2.12 & 2.77 & \phantom{}14.87 & 0.70 & 1.82 \\
E20HLW00LS220 & 3.13 & \phantom{0}18.15 &  0.242 & \phantom{}17.63 & 1.666 & 2.41 & 2.25 & 3.14 & \phantom{}14.79 & 0.70 & 1.66 \\
E25HLW00LS180 & 1.83 & \phantom{00}8.05 &  0.089 & \phantom{0}5.92 & 1.103 & 2.23 & 2.08 & 1.76 & \phantom{0}6.44 & 0.46 & 1.97 \\
E25HLW00LS220 & 1.83 & \phantom{00}8.05 &  0.088 & \phantom{0}5.90 & 1.703 & 2.37 & 2.18 & 1.83 & \phantom{0}5.76 & 0.44 & 1.68
\enddata

\tablecomments{The $u$ series of presupernova models in this table are
  taken from the $Z_\odot = 10^{-4}$ model set of \cite{whw:02}.  We
  imposed a rotation law via equation \ref{eq:specificj}, the value of
  $j_{16,\ \infty}$ is given in the model name following the letter J.
  The m35OC presupernova model is taken from \cite{woosley:06} and
  both the E20 and E25 models are from \cite{heger:00}, these model
  are evolved with rotation.  In simulations where a BH did not form
  within $3.5\,$s we give, in parenthesis, the values at this time.}

\tablenotetext{a}{Initial central angular velocity of star.}
\tablenotetext{b}{Total angular momentum of star.}
\tablenotetext{c}{Initial $T/|W|$ of the star.}
\tablenotetext{d}{$T/|W|$ of the star at bounce.}
\tablenotetext{e}{Angular momentum of protoblack hole (PBH) when
  $\alpha_c=0.3$.}
\tablenotetext{f}{$T/|W|$ of the star when $\alpha_c = 0.3$.}
\tablenotetext{g}{Dimensionless spin of the PBH when $\alpha_c = 0.3$.
  Unphysical values for a BH are shown in braces [$\cdots$].}
\tablenotetext{h}{Total neutrino luminosity averaged over postbounce
  time.}

\label{tab:rotatingresults}
\end{deluxetable*}

\subsubsection{Rotating Progenitors and the Connection to long GRBs} 
\label{sec:grb}

The rotation law of Equation~(\ref{eq:specificj}) qualitatively follows
the predicted angular velocity distribution in the inner $\sim
1-3\,M_\odot$ of presupernova models evolved with rotation. However,
Equation~(\ref{eq:specificj}) asymptotes to constant specific angular
momentum $j$ and cannot capture jumps and secular increase of $j$ in
overlying mass shells (e.g.,~\citealt{heger:00}).

Here we consider three different supernova progenitors evolved with
rotation that have the potential of forming BHs soon after bounce. Models
E20 and E25 are rapidly rotating unmagnetized solar-metallicity
presupernova models of a 20 and 25 $M_\odot$ ZAMS stars from
\cite{heger:00}. Model m35OC of \cite{woosley:06} with 
$M_{\mathrm{ZAMS}} = 35\,M_{\odot}$ has $10\,\%$ solar-metallicity, reduced
mass loss, and magnetic fields. These presupernova models have
initial central angular velocities of $\sim 3.13$, $\sim 1.83$, and
$\sim 1.98\,$rad$\,$s$^{-1}$ and values of $\xi_{2.5}$ of $\sim 0.319$,
$\sim 0.294$, and $\sim 0.456$ for the E20, E25, and m35OC models,
respectively. Due to their moderate $\xi_{2.5}$, we perform
collapse simulations of models E20 and E25 with the LS180 and LS220
EOS. Model m35OC is calculated with all four EOS. 
The progenitor characteristics are summarized in
Tables~\ref{tab:initialmodels} and \ref{tab:rotatingresults}.

For a given EOS, due to very similar $\xi_{2.5.}$, the evolutions of
E20 and E25 are alike. They form BHs in $\sim 1.1~\,$s with the LS180 EOS
and in $\sim 1.7\,$s with the LS220 EOS. Model E20 is more rapidly
spinning. Its $T/|W|$ peaks at $\sim 0.18$ at bounce and settles down
to a nearly constant value of $\sim 0.15$ throughout the postbounce
evolution. It's $a^\star_\mathrm{PBH}$ is $\sim 0.7$ for both the LS180 and
LS220 EOS. Model E25 reaches $T/|W| \sim 0.06$ at bounce and $\sim
0.065$ at BH formation with $a^\star_\mathrm{PBH} \sim 0.45$ for both EOS.

The core of the m35OC model is sufficiently compact to form a BH soon
after bounce if no explosion is launched (e.g., via magneto-rotational
explosion; \citealt{dessart:08a}).  The nascent BH forms at a time of
$\sim 0.75\,$s, $\sim 0.97\,$s, $\sim 2.19\,$s, and $\sim1.91\,$s for
the LS180, LS220, LS375, and HShen EOS, respectively. The initial
gravitational (baryonic) BH masses are $\sim 2.22$ ($\sim
2.33$)$\,M_\odot$, $\sim 2.29$ ($\sim 2.43$)$\,M_\odot$, $\sim 2.69$
($\sim 3.01$)$\,M_\odot$, and $\sim 2.60$ ($\sim 2.84$)$\,M_\odot$ for
the LS180, LS220, LS375, and HShen EOS, respectively. The BHs are
modestly-rapidly spinning with $a^*_\mathrm{PBH}$ of $\sim 0.61$,
$\sim 0.58$, $\sim 0.54$, and $\sim 0.55$ for the LS180, LS220, LS375
and HShen EOS, respectively. For all EOS, the PNS, during the
accretion phase, has a modest $T/|W|$ of $\lesssim 0.12$.

Once a BH is formed, material from the stellar mantle will continue to
accrete at high rates. Accretion will only be slowed once material
with sufficiently high specific angular momentum reaches small radii
and becomes centrifugally supported, forming an accretion disk. This
is the crucial prerequisite for the collapsar scenario for long GRB
central engines to work \citep{woosley:93}. Models E20 and E25 lost
much of their initial mass and angular momentum during their evolution
to the presupernova stage and, therefore, there is too little angular
momentum in their outer regions to allow for a long-term accretion
disk.

\begin{figure}[t]
\begin{center}
\includegraphics[width=\columnwidth]{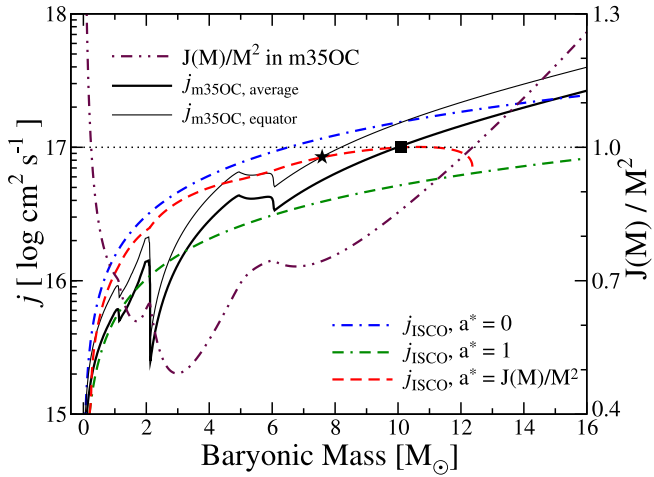} 
\caption{Specific angular momentum (solid, left ordinate) and
  dimensionless spin (dot-dot-dashed, right ordinate) for the GRB
  progenitor model m35OC from \cite{woosley:06} as a function of
  enclosed baryonic mass.  The thin solid line is the angular momentum
  at the equator; the thick solid line is the angle-averaged angular
  momentum.  Also shown, dash-dotted, dash-dash-dotted,
  and dashed, is the specific angular momentum for a mass
  element at the innermost stable orbit assuming a background
  spacetime being Schwarzschild, maximally Kerr, and Kerr with $a^* =
  J(M)/M^2$, respectively.  The horizontal line denotes $a^* =
  1$.  ($\bigstar$) and ($\blacksquare$) denote the mass
  coordinate where the equatorial and angle-average specific angular
  momenta exceed $j_\mathrm{ISCO}$, respectively.\label{fig:m35specificj}}
\end{center}
\end{figure}

The situation is different for the m35OC progenitor.  Its particular
evolution prevented dramatic loss of mass and angular momentum while
keeping its envelope radius small. In Figure~\ref{fig:m35specificj},
we show the specific angular momentum distribution of the m35OC
progenitor as a function of enclosed baryonic mass. We also include
graphs of the $j_\mathrm{ISCO}$, the specific angular momentum
required for a stable orbit at the innermost stable circular orbit
(ISCO) of a Kerr hole with mass $M$ and spin $a^\star$
\citep{bardeen:72}. Curves for $a^*=0$ (Schwarzschild), $a^*=1$
(maximally Kerr), and $a^*$ = $J(M)/M^2$ are shown.  We also plot the
value of $a^*$ that a BH of baryonic mass $M$ formed from the m35OC
progenitor would have. Figure~\ref{fig:m35specificj} is independent of
the detailed collapse evolution, assuming that no angular momentum is
radiated by neutrinos and/or gravitational waves, or ejected. However,
we note that due to emission of neutrinos before BH formation, the
enclosed gravitational mass (entering into the calculation of $a^*$)
will be smaller by up to $\sim 0.2 - 0.4\,M_\odot$ than the baryonic
mass given in the figure. This leads to slightly underpredicted values
of $a^*$ for small $M$. Since relatively little energy is emitted in
neutrinos after BH formation, the relative discrepancy between
gravitational and baryonic mass decreases with growing BH mass.

Figure~\ref{fig:m35specificj} can be interpreted as follows. If the
CCSN mechanism fails to reenergize the shock and the PNS collapses to
a BH of mass $M$, then its initial angular momentum $J$ and spin
$a^\star$ will be set by the enclosed angular momentum and
gravitational mass. Initially, hyperaccretion will increase both $J$
and $M$, but $a^\star$ may increase or decrease, depending on the
angular momentum of the accreted matter. Accretion will slow down and
a disk will form once infalling material has specific angular momentum
$j$ greater than $j_\mathrm{ISCO}$. In model m35OC, this occurs
between a BH mass coordinate of $\sim 7.6\,M_\odot$ (for a mass
element with equatorial $j$) and $\sim 10.1\,M_\odot$ (for a mass
element with angle-averaged $j$). These points are marked in
Figure~\ref{fig:m35specificj} with a ($\bigstar$) and
($\blacksquare$), respectively. Using accretion simulations with
\code{GR1D} setup to include an inflow inner boundary condition we
find that with the m35OC model, the accretion time for $7.6\,M_\odot$
and $10.1\,M_\odot$ is $\sim 10.2\,$s and $\sim 14.3\,$s from the
onset of collapse, respectively. These times are roughly twice the
free fall time, since material in outer regions in hydrostatic
equilibrium for a sound travel time \citep{burrows:86bh}. Once the
disk has formed, accretion will continue via processes that will
transport angular momentum out and mass in. A collapsar central engine
may begin its operation \citep{woosley:93,macfadyen:99} with a central
BH of $M \sim 8\,M_\odot$, $a^* \sim 0.75$, and an ISCO radius of
$\sim 40\,$km.

\begin{figure*}[t]
\begin{center}
\includegraphics[width=1.5\columnwidth]{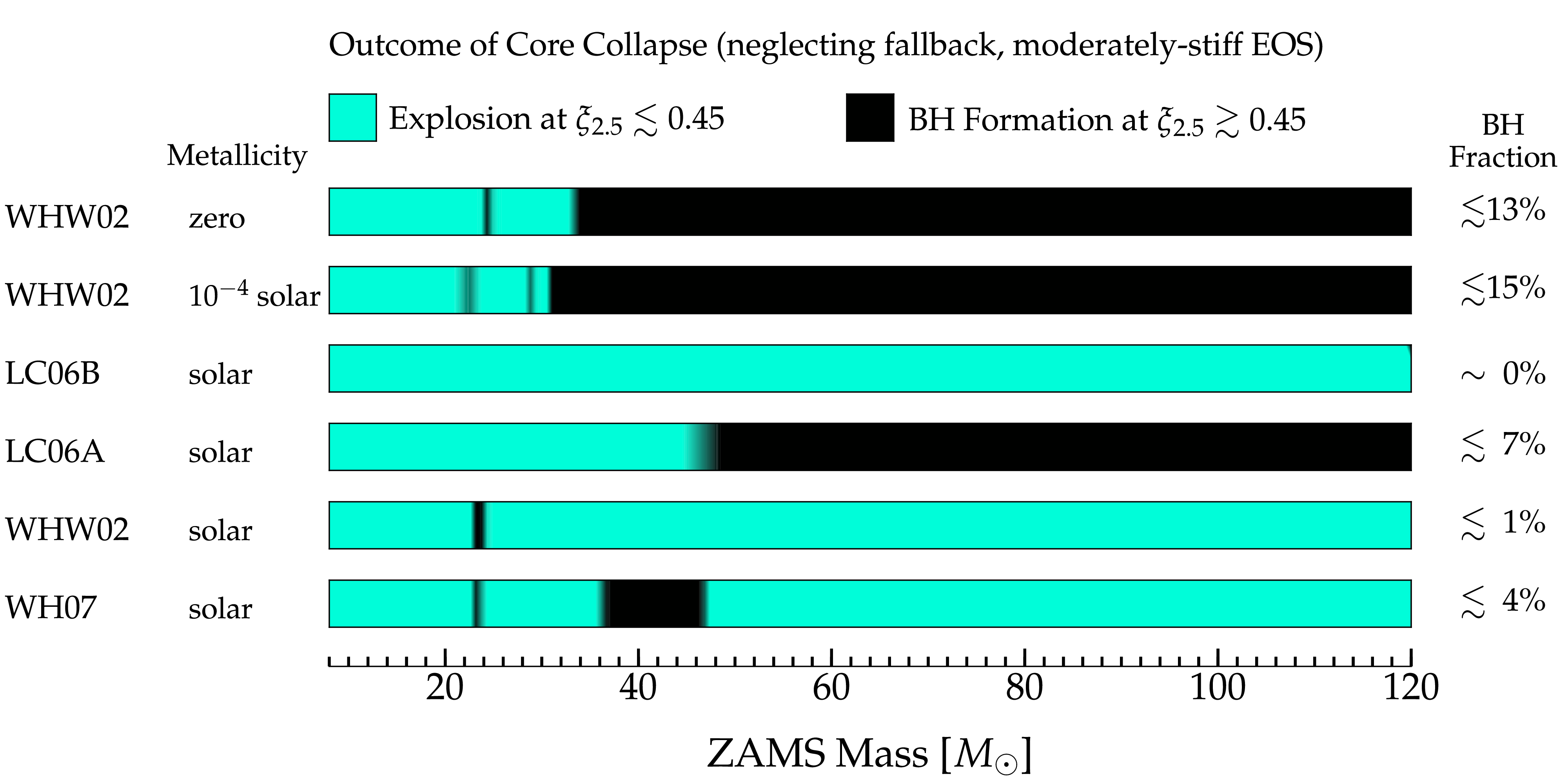} 
\caption{Outcome of core collapse as a function of ZAMS mass of single
  nonrotating massive stars, assuming that for moderately stiff
  nuclear EOS (e.g., LS180/LS220), neutrino-driven explosions can be
  launched up to a bounce compactness $\xi_{2.5} \lesssim 0.45$
  (cf.~Section \ref{sec:lums}). Other potential explosion mechanisms are
  neglected.  We consider only explosion and BH formation
    without explosion as outcomes and neglect other scenarios,
  including post-explosion BH formation via fallback accretion
  \citep{zhang:08,dessart:10}, cooling or nuclear phase transitions.  Shown are
  results for a range of model sets and metallicities (see
  Section \ref{sec:initialmodels}).  Very low metallicity stars with ZAMS
  masses above $\sim 30\,M_\odot$ robustly form a BH without
  explosion. At higher metallicity, uncertainties in the physics of
  mass loss (e.g., \citealt{smith:10}) make robust predictions
  difficult. This is reflected in the rather dramatic disagreement of
  the four solar-metallicity progenitor model sets that we include.
  The ``BH fractions'' stated at the right edge of the plot denote the
  fraction of massive stars with $M\gtrsim 8\, M_\odot$ that form
  BHs. They are obtained by convolution with a Salpeter IMF under the
  assumption that stars with $8\,M_\odot \lesssim M \lesssim
  14\,M_\odot$ explode robustly.}
\label{fig:outcome}
\end{center}
\end{figure*}

\section{Summary and Conclusions}
\label{sec:conclusion}

We have performed an extensive study of BH formation in failing
CCSNe with the open-source 1.5D GR code \code{GR1D},
making the simplifying assumptions of spherical symmetry and of a
neutrino leakage scheme rather than full Boltzmann transport.  We have
performed more than 700 collapse simulations with over 100 unique
progenitor models, probing systematically the many-dimensional
parameter space that determines the outcome of stellar collapse in
single massive stars. Specifically, we have studied and established
the systematic dependence of CCSN failure and BH formation on
progenitor compactness, precollapse rotational setup, neutrino
heating efficiency, and nuclear EOS.

To first approximation, the evolution of any core collapse event
proceeds as follows.  Core collapse ensues in a given presupernova
star, collapse is halted when the inner core of $\sim 0.5 -
0.7\,M_\odot$ reaches nuclear density. A shock is formed, propagates
outward initially in $M$ and $r$, but soon stalls. Assuming the CCSN
mechanism, whatever its precise nature may be, fails, we can robustly
predict the time it takes to BH formation for a given nuclear EOS
(scaling with EOS stiffness) based on a single parameter, the
progenitor bounce compactness $\xi_{2.5}$ ($t_\mathrm{BH} \propto
{\xi_{2.5}}^{-3/2}$). Using the same parameter, for a given EOS, we
can predict the maximum mass of the PNS at collapse and its thermal
enhancement ($10$\%-$25\%$) over the cold NS mass, due, as we have shown
for the first time by detailed comparison with exact TOV solutions,
primarily to thermal pressure support in the outer PNS core.

In an attempt to more quantitatively understand which stars explode
and which do not, assuming the neutrino mechanism is responsible for
the majority of CCSN explosions, we have turned the knobs on
\code{GR1D}'s neutrino heating scheme, experimentally, to first order,
establishing the neutrino heating efficiency needed to explode a
progenitor with given $\xi_{2.5}$. Neglecting the potentially highly
relevant effects of multi-dimensional dynamics and assuming an EOS of
intermediate stiffness (the LS220 EOS), we predict that progenitors
with bounce compactness $\xi_{2.5} \gtrsim 0.45$ most likely form BHs
without explosion. This prediction, in itself, without connection to
ZAMS conditions through stellar evolution, is of limited
utility. Using the whole set of progenitor data made available to us
by stellar evolution groups, we attempt the former in
Figure~\ref{fig:outcome}. We plot the mapping between ZAMS mass and
outcome of core collapse, reduced to explosion or no explosion and BH
formation, neglecting completely the possibility of BH formation due
to fallback/cooling/phase transitions after a launched explosion.  The
case is clear cut at low metallicity where mass loss has negligible
effect on the mapping between ZAMS conditions and core collapse
outcome.  Using a Salpeter initial mass function (IMF; $\alpha =
2.35$, $M_\mathrm{min} = 8.0\,M_\odot$, and $M_\mathrm{max} =
150.0\,M_\odot$) we estimate that $\sim$15\% of all progenitors form
BHs without explosion.  At (around) solar metallicity, the precise way
of prescribing mass loss in stellar evolution has tremendous
consequences on the mapping between ZAMS mass and core collapse
outcome. Depending on the particular mass-loss prescription, we
predict a BH fraction of $0\%$-$7\%$ for solar-metallicity
stars. This makes mass loss the single most important unknown
parameter in connecting ZAMS conditions to core collapse outcome (in
agreement with \citealt{smith:10}).

Rapid rotation, which may be present in a significant subset of
massive stars, generally increases the maximum PNS mass by centrifugal
support and delays BH formation. Assuming (quite likely) uniform
rotation of the PNS core, the increase in maximum PNS mass due to
centrifugal support in the range of rotation rates explored is $\sim
5$\%-$10\%$. In the basic neutrino mechanism, rotation leads to a lower
sum of $\nu_e$ and $\bar{\nu}_e$ luminosities and lower mean energies
for all neutrino types. This is detrimental for explosion in 1.5D (and
perhaps even in 2.5D) despite centrifugal support
\citep{fh:00,ott:08}. A larger fraction of massive stars may form BHs
with (moderate) rotation than without. Left out of this picture are
potential magnetohydrodynamics contributions to the explosion
mechanism and energetics (cf.~\citealt{burrows:07b}).

Of particular interest to both formal relativity theory and
astrophysics is the range of potential birth spins of BHs. Our results
quite strikingly suggest that the rotation rate of the maximum-mass
PNS and, hence, the spin of the nascent BH, will be limited to values
of $a^\star$ below $\lesssim 0.9$ by likely nonaxisymmetric dynamics.
If true and confirmed by multi-dimensional simulations, 3D rotational 
instabilities may be a cosmic censor preventing naked
singularities from forming in stellar collapse. 

Rotation and the associated angular momentum are key ingredients in
the collapsar scenario for GRBs~\citep{woosley:93}. As part of this
study, we have performed the first BH formation study with the m35OC
GRB progenitor of \cite{woosley:06}. Using the LS220 EOS, we predict
an initial BH mass of $\sim 2.29\,M_\odot$ and $a^\star$ of $\sim
0.58$. Assuming that the GRB engine cannot operate until a Keplerian
disk has formed, there will be a delay of $\sim 10\,\mathrm{s}$
between BH formation and GRB engine ignition at a BH mass of $\sim 8\,
M_\odot$ and $a^\star \sim 0.75$.

Finally, we re-emphasize that the goal of this study was not to yield
accurate predictions about the outcome of core collapse in any
individual progenitor.  Rather, we have studied and established
overall trends with progenitor parameters. We have made
simplifications and approximations, and have omitted a broad range of
potentially relevant physics. The most important of the latter may
well be multi-dimensional dynamics and their effect on the CCSN explosion
mechanism and on the associated failure rate of CCSNe.

Future work may be directed toward studying the systematics of BH
formation in the post-explosion phase via fallback accretion, PNS
cooling, or EOS phase transitions. Our current neutrino treatment must
be upgraded for more quantitatively accurate simulations and neutrino
signature predictions. Ultimately, multi-dimensional GR simulations 
of successful and failing CCSNe will be necessary to study
the multi-dimensional dynamics left out here and for making truly robust
predictions of the outcome of stellar collapse for any given
set of initial conditions.

We acknowledge helpful discussions with and input from A.~Burrows,
P.~Cerd\'a-Dur\'an, L.~Dessart, M.~Duez, T.~Fischer, J.~Kaplan,
J.~Lattimer, C.~Meakin, J.~Murphy, F.~Peng, S.~Phinney, C.~Reisswig,
S.~Scheidegger, N.~Smith, E.~Schnetter, K.~Thorne, and S.~Teukolsky.
We thank S.~Woosley and A.~Heger for their recent presupernova models
and A.~Chieffi and M.~Limongi for making available both of their
presupernova model sets.  The computations were performed at Caltech's
Center for Advanced Computing Research on the cluster ``Zwicky''
funded through NSF grant no.\ PHY-0960291 and the Sherman Fairchild
Foundation. Furthermore, computations were performed on Louisiana
Optical Network Infrastructure computer systems under allocation
loni\_numrel05, on the NSF Teragrid under allocation TG-PHY100033, and
on resources of the National Energy Research Scientific Computing
Center, which is supported by the Office of Science of the
U.S. Department of Energy under Contract No. DE-AC02-05CH11231.  This
research is supported in part by the National Science Foundation under
grant nos. AST-0855535 and OCI-0905046.  EOC is supported in part by a
post-graduate fellowship from the Natural Sciences and Engineering
Research Council of Canada (NSERC).

\end{document}